\documentclass[trackchanges]{aastex701}

\usepackage{amsmath}
\usepackage[T1]{fontenc}
\usepackage{appendix}

\begin{document}

\title{Numerical Simulations of the interaction between winds from tidal disruption event and supermassive black hole circumnuclear medium}

\author[orcid=0000-0002-0427-520X,gname=De-Fu, sname='Bu']{De-Fu Bu} 
\affiliation{Shanghai Key Lab for Astrophysics, Shanghai Normal University, 100 Guilin Road, Shanghai 200234, China}
\email[show]{De-Fu Bu (dfbu@shnu.edu.cn)}

\author[orcid=0009-0008-9058-7023,gname=Yu-Heng,sname='Sheng']{Yu-Heng Sheng}
\affiliation{Astrophysics Division, Shanghai Astronomical Observatory, Chinese Academy of Sciences, 80 Nandan Road, Shanghai 200030, People’s Republic of China}
\affiliation{School of Astronomy and Space Sciences, University of Chinese Academy of Sciences, 19A Yuquan Road, Beijing 100049, China}
\email[show]{shengyuheng@shao.ac.cn}

\author[orcid=0000-0002-1908-0536, gname=Liang, sname='Chen']{Liang Chen}
\affiliation{Astrophysics Division, Shanghai Astronomical Observatory, Chinese Academy of Sciences, 80 Nandan Road, Shanghai 200030, People’s Republic of China}
\email[]{chenliang@shao.ac.cn} 

\author[orcid=0000-0002-0092-7944, gname=Guobin, sname='Mou']{Guobin Mou}
\affiliation{Department of Physics and Institute of Theoretical Physics, Nanjing normal university, Nanjing 210023, People’s Republic of China}
\email[]{} 

\author[orcid=0009-0004-3881-674X,gname=Haojie, sname='Xia']{Haojie Xia} 
\affiliation{Astrophysics Division, Shanghai Astronomical Observatory, Chinese Academy of Sciences, 80 Nandan Road, Shanghai 200030, People’s Republic of China}
\affiliation{School of Astronomy and Space Sciences, University of Chinese Academy of Sciences, 19A Yuquan Road, Beijing 100049, China}
\email[]{}

\author[orcid=0000-0002-2419-9590,gname=Xiao-Hong, sname='Yang']{Xiao-Hong Yang} 
\affiliation{Department of Physics, Chongqing University, Chongqing 400044, China }
\email[]{yangxh@cqu.edu.cn}

\begin{abstract}
The interaction between winds from tidal disruption event (TDE) and the circumnuclear medium (CNM) around supermassive black holes is believed to contribute to the radio emission of non-jetted TDEs. This model has been studied both analytically and numerically. In previous works, the properties of wind are usually set to be angle-independent. It is not clear to what extent the results deviate from those in the case where TDE wind properties are angle-dependent. We perform numerical simulations to investigate the effects of the angular distribution of wind properties on the radio light curve in the wind-CNM interaction model. We find that given that the wind mass and kinetic energy fixed, the radio light curve in the optically thin waveband significantly depends on the angle-dependence of wind properties. Even when the swept-up CNM mass is still less than the wind mass, a slight deceleration can affect the light curve. Changing the angular dependence of the wind properties causes the deceleration to vary across angles. Consequently, the final light curve—the time evolution of radiation integrated over all angles—would differ significantly from that predicted in models with a spatially uniform wind injection. This can result in a slower rise or a faster decline (depending on the CNM profile) than the analytical prediction, and can even turn the rise phase into a long-lasting plateau. Therefore, it is essential to carefully take into account the angular dependence of wind properties to study the wind-CNM interaction.

\end{abstract}



\section{INTRODUCTION}

When a star passes too close to a central supermassive black hole (SMBH), such that its pericenter distance ($R_{\rm p}$) falls below or equals the tidal disruption radius ($R_{\rm T}$) (\cite{1975Natur.254..295H}; \cite{Rees1988}; \cite{Evans1989}), the star gets disrupted by the SMBH's tidal force, giving rise to a tidal disruption event. While roughly half of the stellar debris gets ejected, the other half remains gravitationally bound and later falls back toward the vicinity of the SMBH. The fallback rate evolves with time as $\dot{M}_{\rm fb} \propto t^{-5/3}$ (\cite{Evans1989}; \cite{Phinney1989}; \cite{Lodatoetal2009}; \cite{GuillochonRamirezRuiz2013}). The peak debris fallback rate can exceed the Eddington accretion rate for a typical TDE, in which a solar-type star is disrupted by an SMBH of $10^{6}-10^{7}\,M_{\odot}$, where $M_{\odot}$ represents the solar mass. Such a high mass fallback rate is sufficient to power a super-Eddington accretion flow. For a typical TDE, the accretion flow is expected to have a size of $\sim 10^{13}~{\rm cm}$ and a temperature of $\sim 10^{5}~$K (see \cite{2020SSRv..216..114R} for a review).

The emission spectrum of such a system is expected to peak in the soft X-ray band (\cite{Cannizzoetal.1990}; \cite{Ulmer1999}; see \cite{2020SSRv..216..114R} for reviews), and indeed several TDEs have been detected in this band (see \cite{Komossa2015}; \cite{2020SSRv..216...85S} for a review). Over the past few decades, however, a large number of TDEs have been discovered in the optical/UV bands. These events exhibit inferred radiation sizes ($\sim10^{14}-10^{16}$~cm) that are significantly larger than theoretical expectations (\cite{Hungetal2017}; \cite{vanVelzenet42al.2020}; \cite{Gezari2021}), and their radiation temperatures are much lower ($\sim\!10^{4}$~K; e.g., \cite{Gezarietal.2012}; \cite{Arcavietal.2014)}). Two main scenarios have been proposed to account for this discrepancy. The first scenario invokes shocks generated by collisions among debris streams during the fallback phase (\cite{Guillochon&Ramirez-Ruiz2015}; \cite{Shiokawaetal.2015}), which dissipate kinetic energy and power the observed optical/UV emission (\cite{Piranetal.2015}; \cite{JiangGuillochonLoeb2016}; \cite{Steinberg&Stone2024}; \cite{2025ApJ...979..235G}). The second scenario posits that the accretion flow in the super-Eddington phase develops an optically thick wind/envelope which envelopes the small accretion disk. The presence of strong winds in TDEs have recently been verified by observations (e.g., \cite{2017ApJ...846..150Y}; \cite{2018MNRAS.474.3593K}; \cite{2020MNRAS.494.4914P}; \cite{2025ApJ...989L...9L}; \cite{2019ApJ...879..119H}; \cite{2024ApJ...972..106X}) and numerical simulations (\cite{Curd&Narayan2019}; \cite{Dai2018}; \cite{Thomsen2022}; \cite{BU22}; \cite{BU23};\cite{2026ApJ..1001..192S}; \cite{2026ApJ...998...23S}; \cite{2026A&A...705A...4B}; \cite{2026ApJ..1004...37L}). High-energy photons generated near the SMBH are then reprocessed into optical/UV photons as they propagate through the expanding optically thick envelope (\cite{Loeb&Ulmer1997}; \cite{Coughlin&Begelman2014}; \cite{Rothetal.2016}; \cite{Liuetal.2017}; \cite{Liuetal.2021}; \cite{Metzger&Stone2017}; \cite{Metzger2022}; \cite{Weversetal.2022}) or wind (\cite{2009MNRAS.400.2070S}; \cite{Lodato&Rossi2011}; \cite{Metzger&Stone2016}; \cite{Piro&Lu2020}; \cite{Uno&Maeda2020}; \cite{BU22}; \cite{Parkinsonetal.2022}; \cite{MageshwaranShawBhattacharyya2023}; \cite{Curd&Narayan2019}; \cite{Dai2018}; \cite{Thomsen2022}; \cite{2026ApJ..1001..192S}).

In additon to optical/UV and X-ray emissions, radio emissions are detected in over 50 TDEs (\cite{2026arXiv260306056Z}). These sources include several jetted-TDEs, including Sw J1644+57 (e.g., \cite{2011Sci...333..203B}; \cite{2011Sci...333..199L}; \cite{2011Natur.476..425Z}; \cite{2012MNRAS.420.3528M}), Sw J2058+05 (\cite{2012ApJ...753...77C}), Sw J1112-82 (\cite{2015MNRAS.452.4297B}), AT2022cmc (\cite{2022Natur.612..430A}) and EP250702a (e.g., \cite{2026SciBu..71..538L}). For the jetted TDEs, the radio luminosity is higher than $10^{40} {\rm erg/s}$. 
The radio emission from the non-jetted TDEs has much lower radio lumininosity $< 10^{40} {\rm erg/s}.$ According to the radio flares time relative to the optical peak time, these TDEs can be divided into two types, a prompt flare that appears within weeks of the optical peak (e.g., \cite{2016ApJ...819L..25A}; \cite{2021ApJ...919..127C};
\cite{2022MNRAS.511.5328G}; \cite{2023MNRAS.522.5084G}; \cite{2023MNRAS.518..847G}; \cite{2024MNRAS.528.7123G}; \cite{2024ApJ...973..104D}), and a delayed flare that emerges months to years after the optical discovery (\cite{2020SSRv..216...81A, 2021NatAs...5..491H, 2021ApJ...920L...5H, 2022ApJ...938...28C, 2024ATel16650....1C, 2022ApJ...925..143P, 2022ApJ...933..176S, 2024ApJ...962L..18Z, 2023MNRAS.520.2417W, 2025SciA...11y9068W}).

The radio emission is generally attributed to synchrotron radiation from cosmic-ray electrons (CRe) accelerated at forward shocks (FS) generated by the interaction of jets or winds with CNM (\cite{2013ApJ...772...78B, 2021MNRAS.507.4196M, 2024ApJ...971...49M, 2024ApJ...971..185C, 2026ApJ..1006...70M, 2025ApJ...988L..24H, 2025arXiv251223252S, 2026ApJ..1007..103M, 2023ApJ...954....5H, 2026ApJ...998..199W,2026ApJ..1006..103L}; \cite{2023ApJ...957L...9T}; \cite{2020MNRAS.492..686L}; \cite{christy2025dichotomylonglivedradioemission}; \cite{2025ApJ...993..207B}; \cite{2023MNRAS.520.2417W}; \cite{Farley_2026}; \cite{mummery2026tripleradioflarestidal})). Other studies have suggested that CRe may instead be accelerated in bow shocks produced by the interaction between outflows and dense clouds in the vicinity (\cite{Mou2022, 2023MNRAS.521.4180B, 2025ApJ...979..109Z, 2025arXiv251025033M, 2025ApJ...993L...2Y}). Additional studies have proposed that CRe can originate from shocks arising from the interaction of unbound debris with the CNM (\cite{2016ApJ...822...48G}; \cite{2016ApJ...827..127K, 2019MNRAS.487.4083Y};  \cite{2022MNRAS.515.1699S}) or with a torus (\cite{2024ApJ...977...63L}).

While the exact origin of these CRe remains unclear, a forward shock is expected to form whenever a fast wind is launched and propagates through the CNM. Analytic treatments of the radio emission from wind-CNM interaction have been developed by \cite{2013ApJ...772...78B} and \cite{2021MNRAS.507.4196M}. Hydrodynamical simulations have also been carried out to compute the radio properties of this interaction. In some simulations, the CRe is not included. In the post-processing process, the CRe is included according to the shock conditions to calculate the synchrotron emission (e.g., \cite{2025ApJ...988L..24H}; \cite{2026ApJ..1006..103L}). In contrast, \cite{2026ApJ..1006...70M} perform simulations that directly inject CRe in the simulations and tracing the evolution of the CRe, which allows them to self-consistently obtain the spatial distribution of CRe and compute the radio emission more precisely.

Nevertheless, to the best of our knowledge, all of the aforementioned studies adopt a prescribed wind injection scheme, in which the wind is continuously injected into the simulation domain with properties that are both time-independent and angular-uniformly distributed. However, recent hydrodynamical simulations have shown that the wind launched in a TDE exhibits both angular dependence and temporal evolution (e.g., \cite{Dai2018,BU23,2026ApJ...998...23S}). It is not clear how does the time-dependent and angular non-uniform nature of wind affect the radio emissions. Therefore, it is very important to re-study the wind-CNM interaction with wind properties from numerical simulations and compare to the commonly adopted time-independent and angular-uniform wind-CNM interaction models. 

In our previous work (\cite{BU22}), we performed two-dimensional radiative hydrodynamical simulations to investigate the temporal evolution of the circularized accretion system for typical TDE systems, in which a solar-mass star is disrupted by an SMBH of $10^6~M_{\odot}$ and by an SMBH of $10^7~M_{\odot}$, respectively. In a subsequent study (\cite{BU23}), we examined the temporal evolution and spatial distribution of the wind in each system. Accordingly, we have obtained a more realistic description of TDE winds than the uniform, time-independent prescription adopted by previous works. 

In this work, we carry out hydrodynamical simulations to study the wind-CNM interaction. The wind properties are from those obtained in \cite{BU23}. In our simulations, CRe are injected according to the shock conditions. We trace the time-evolution of the CRes analogous to that of \cite{2025arXiv251025033M}. Based on the simulated CRes distribution, we calculate the synchrotron emission and then generate the radio-band spectra and light curves. By comparing these results with those from simulations employing uniform wind injection, we investigate the effects of the wind properties on the radio emission.

The structure of this paper is as follows. In Section 2, we describe the simulation setups and the methodology adopted in our calculations. Our results are presented in Section 3. We summarize and discuss our findings in Section 4.

\section{METHOD}

We perform 2D axisymmetric hydrodynamic simulations using the PLUTO code (\cite{2007ApJS..170..228M}) in the spherical coordinates ($r$, $\theta$,$\phi$), incorporating the wind injection module and the CRe injection module that we describe in this section. The simulation domain extends radially from $0.0061$ to $2.40$~pc. In the radial direction, we employ 1024 uniformly spaced grids between $0.0061$ and $0.02$~pc, and 6408 logarithmically spaced grids between $0.02$ and $2.4$~pc. In the $\theta$ direction, we use 384 uniformly spaced grids for $0 \le \theta \le \pi/6$ and 384 logarithmically spaced grids for $\pi/6 \le \theta \le \pi/2$. Outflow boundary condition is applied at the inner and outer radial boundaries. At $\theta = 0^\circ$, we adopt the axisymmetric boundary condition, while at $\theta = 90^\circ$, we use the reflecting boundary condition.

\subsection{CNM}

The knowledge of the hot CNM around SMBHs remains limited. Observational constraints on the CNM are available primarily for two sources: Sgr A* and M87. For Sgr A*, the CNM density follows $\rho \simeq 30\,m_{\rm H}\,{\rm cm^{-3}}\,r_{-1}^{-1}$, where $r_{-1} \equiv r/0.1\,{\rm pc}$ (\cite{2006ApJ...640..319X}; \cite{2019ApJ...871..126G}), while for M87, the density at a given radius in parsec is about one order of magnitude higher than that of Sgr A* (\cite{2015MNRAS.451..588R}). In this work, for models with a single power-law CNM density profile, we adopt the CNM profile used in the fiducial model of \cite{2026ApJ..1006...70M}:
\begin{equation}
    \label{eq1}
    \rho(r) = 90\,m_{\rm H}\,{\rm cm^{-3}}\,r_{-1}^{-n},
\end{equation}
where $n$ is the power-law index of the density profile.

We perform some simulations adopting a single power-law CNM density profile. Meanwhile, many studies have suggested that the CNM density profile differs inside and outside the Bondi radius ($R_{\rm B}$) (e.g., \cite{1952MNRAS.112..195B,2004ApJ...613..322Q,2014ApJ...781...82C,2017MNRAS.464.2481G,2019ApJ...871..126G}). Since the CNM density distribution can affect the shock dynamics and consequently the CRe injection, the resulting spectra and light curves may also be influenced by this effect. We therefore carry out an additional set of simulations using a broken power-law CNM density profile:
\begin{equation}
    \label{eq2}
    \rho(r) = 
    \begin{cases}
        90\,m_{\rm H}\,{\rm cm^{-3}} \left(\dfrac{r}{R_{\rm B}}\right)^{-n_i}, & r \le R_{\rm B} \\[6pt]
        90\,m_{\rm H}\,{\rm cm^{-3}} \left(\dfrac{r}{R_{\rm B}}\right)^{-n_o}, & r \ge R_{\rm B}
    \end{cases}
\end{equation}
where $n_i$ and $n_o$ are the power-law indices inside and outside $R_{\rm B}$, set to 2.5 and 0.0, respectively.

In our simulations, we adopt a central SMBH of $10^7\,M_{\odot}$. Assuming a CNM temperature near the Bondi radius of $T_{\rm B} \simeq 3\times10^7$~K, the Bondi radius is then set to $R_{\rm B} = 0.125$~pc.

\subsection{TDE Wind}

In \cite{BU22}, we perform simulations to study the circularized accretion flow in TDEs. We refer to the simulations in \cite{BU22} as `TDE simulation'. In the TDE simulation, we find strong winds. In present simuations, we study the interaction between TDE wind and CNM. We refer the simulations in this paper to `radio simulation'. In order to study the wind-CNM interaction, we employ the wind properties found in the TDE simulation in \cite{BU22}. We extract the wind properties at $r = 0.0061$~pc, which corresponds to the inner boundary of the simulation domain in this paper, and interpolate them onto the same $\theta$ grid used in the present radio simulations. For a TDE in which a $10^7\,M_{\odot}$ SMBH disrupts a solar-mass star, the fallback rate of the debris would exceed the Eddington accretion rate for $\sim 180$~days, during which a strong wind would be launched near the injection location. Considering that the wind needs a few tens of days to travel to $0.0061$~pc, we only compute the wind properties within this 200-day period and inject them at the first radial grid cell of the radio simulation domain. Since the fallback rate then falls below the Eddington accretion rate, the accretion system is expected to transition from the super-Eddington to the sub-Eddington phase. When and how this transition occurs remains an open question, and our simulation resolution is insufficient to resolve the thin disk that forms in the sub-Eddington phase precisely. Therefore, the previous work \cite{BU22} does not extend to later times. Since the wind outflow rate is expected to decrease as the fallback rate declines, for $t > 200$~days we apply an exponential decay to the wind injection. We have performed test simulations with different decay forms and find that they do not significantly affect the final results.

\begin{figure}
    \centering
    \includegraphics[width=0.7\linewidth]{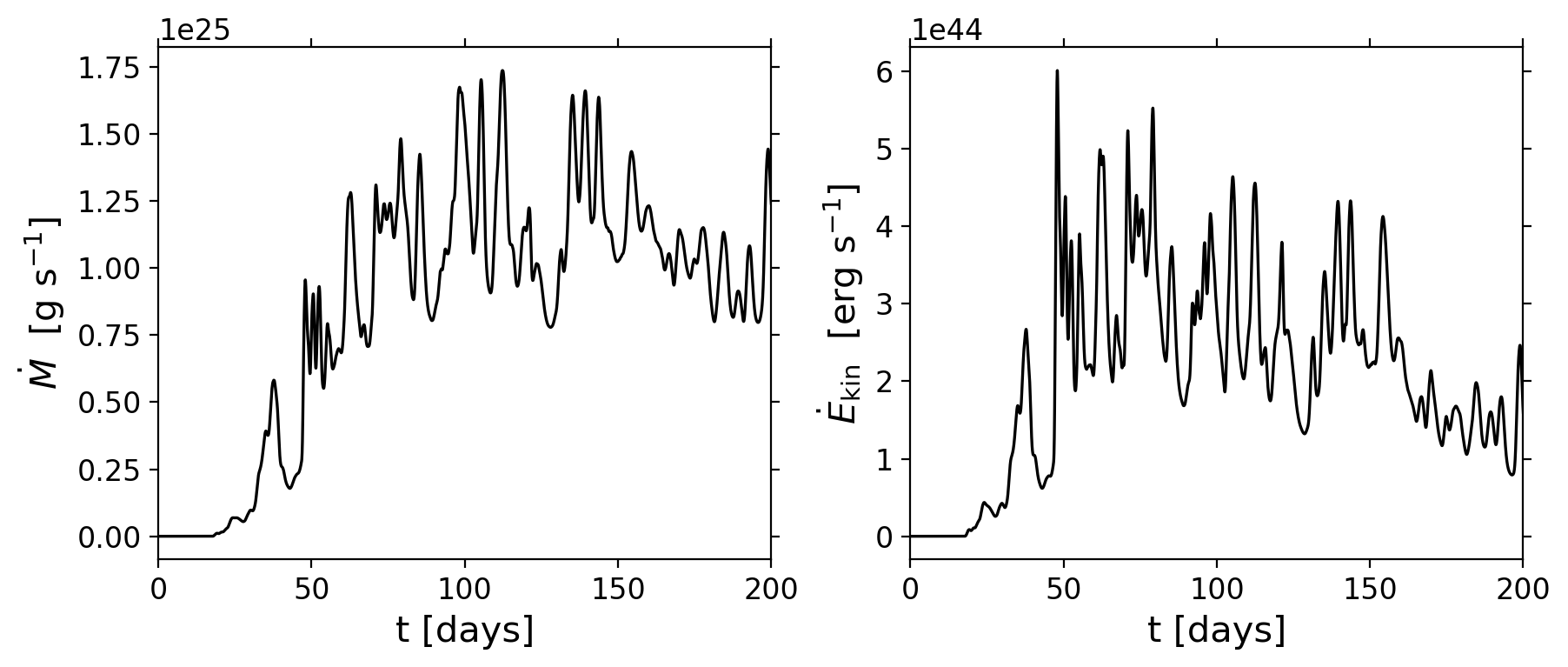}
    \caption{Temporal evolution of the mass outflow rate (left panel) and the kinetic power (right panel) of wind, integrated over the entire hemisphere $0 \le \theta \le \pi/2$. The data are from model M7 of the TDE simulation of \cite{BU22}.}
    \label{fig1}
\end{figure}

\begin{figure}
    \centering
    \includegraphics[width=0.7\linewidth]{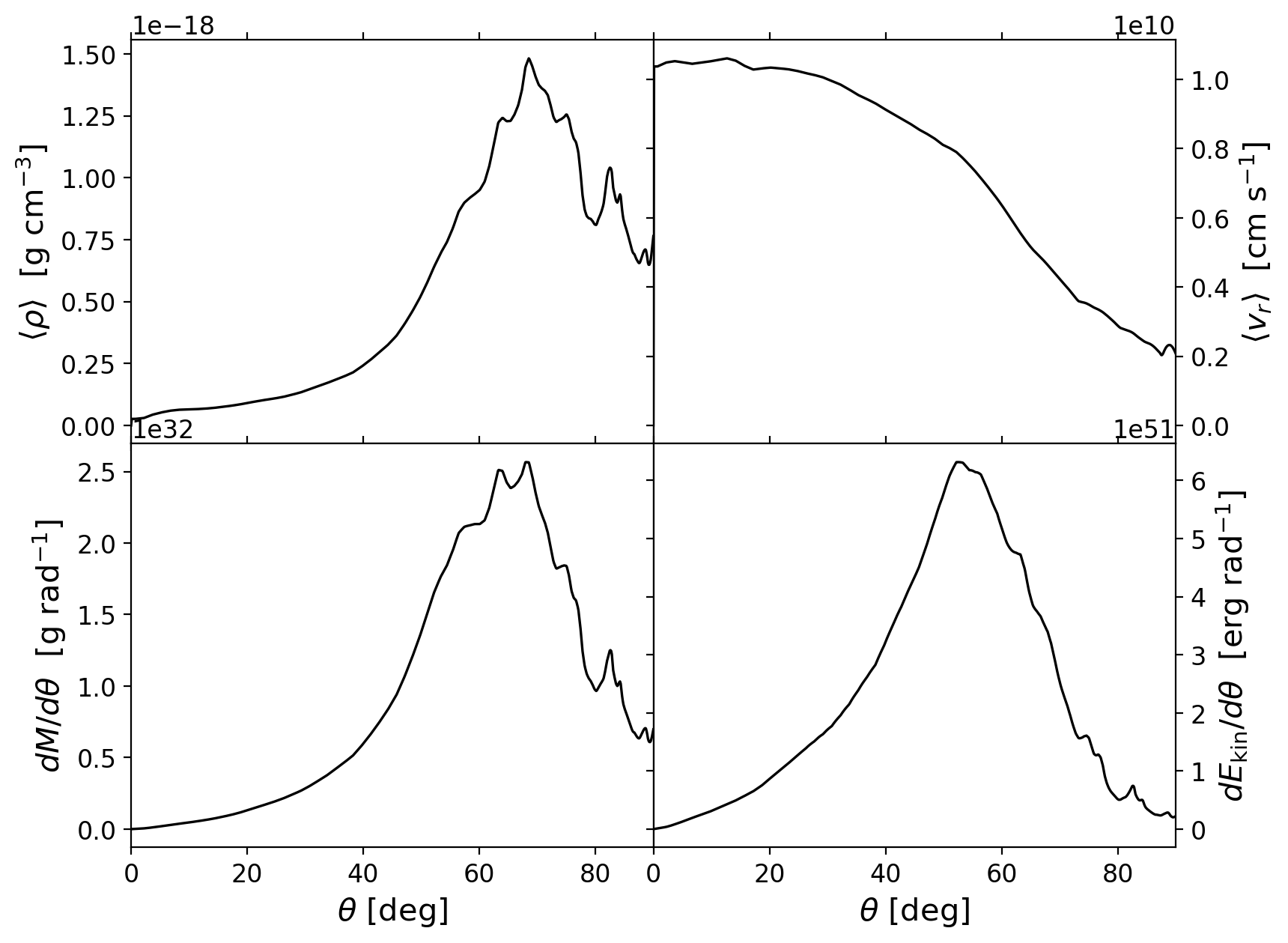}
    \caption{Angular distributions of the time-averaged density (top-left panel), the radial velocity (top-right panel), the total injected mass (bottom-left panel), and the total injected kinetic energy (bottom-right panel), over the 200-day period.}
    \label{fig2}
\end{figure}

Fig.~\ref{fig1} shows the temporal evolution of the mass outflow rate and the kinetic power of the wind. The data are from the model M7 (with a $10^7 M_\odot$ black hole mass) in the TDE simulation of \cite{BU22}. Both the wind mass outflow rate and kinetic power are quite time-variable. However, generally, both the mass outflow rate and the kinetic power of wind increase with time initially and then decline.

Fig.~\ref{fig2} presents the angular distribution of the wind properties averaged over the entire 200-day period. Overall, the radial velocity of the wind decreases with increasing $\theta$, while the density initially increases with $\theta$ and then decreases near the mid-plane region ($65^\circ < \theta < 90^\circ$). Consequently, both the total injected mass and kinetic energy first rise and then fall with increasing $\theta$.

Over the 200-day period, approximately $0.074\,M_{\odot}$ of mass and $2.5\times10^{51}$~erg of kinetic energy are injected into the region $r > 0.0061$~pc and $0 < \theta < \pi/2$. To compare the results with those from uniform wind injections, we perform additional simulations. In some simulations, we inject the wind spherically; while in other simulations, we inject the wind uniformly within a cone of $0 < \theta < 30^\circ$. In these simulations, the properties of the injected wind are time-independent, and the total injected mass and kinetic energy over the 200-day period are the same as those from the simulations employing the wind properties shown in Figure \ref{fig1}. For both the spherical and cone injection models, the wind radial velocity is $5.9\times10^9~{\rm cm\,s^{-1}}$. In some spherical injection models, in order to satisfy that the total injected mass is 0.074$M_\odot$, the injected density is $6.41\times10^{-19}~{\rm g\,cm^{-3}}$; while in the cone injection models, the wind density   is $4.78\times10^{-18}~{\rm g\,cm^{-3}}$. The information of all the models are summarized in Table 1.

\begin{table} \caption{Models}
\setlength{\tabcolsep}{1.mm}{
\begin{tabular}{ccccccccc}
\hline \hline
 Model & CNM profile & ($n$, $n_i$, $n_o$)$^*$ & wind structure  & wind & wind kinetic  & wind velocity & wind density    \\

  &  &  &  & mass ($M_\odot$)  & energy (erg) &  ($ {\rm cm \ s^{-1}} $) & (${\rm g \ cm^{-3}}$) \\

\hline\noalign{\smallskip}
Run-SP-sph  & single power-law & $n=1$ & spherical & $0.074$ & $3.6\times 10^{51} $ & $5.9\times10^9$ & $6.41\times10^{-19}$     \\
Run-SP-sphL  & single power-law & $n=1$ & spherical & $0.005$ & $2.4\times 10^{50} $ & $5.9\times10^9$ & $4.33\times10^{-20}$     \\
Run-SP-cone  & single power-law & $n=1$ & cone & $0.074$ & $3.6\times 10^{51}$ & $5.9\times10^9$ & $4.78\times10^{-18}$   \\
Run-SP-M7   & single power-law & $n=1$ & Figs \ref{fig1} and \ref{fig2} & $0.074$ & $3.6\times 10^{51} $ & Fig \ref{fig2} & Fig \ref{fig2} \\
Run-SP-cone-0  & single power-law & $n=0$ & cone & $0.074$ & $3.6\times 10^{51} $ & $5.9\times10^9$ & $4.78\times10^{-18}$   \\
Run-SP-cone-2.5  & single power-law & $n=2.5$ & cone & $0.074$ & $3.6\times 10^{51} $ & $5.9\times10^9$ & $4.78\times10^{-18}$   \\
Run-BP-sph  & broken power-law & $n_i=2.5, n_o=0$ & spherical & $0.074$ & $3.6\times 10^{51} $ & $5.9\times10^9$ & $6.41\times10^{-19}$    \\
Run-BP-cone  & broken power-law & $n_i=2.5, n_o=0$ & cone & $0.074$ & $3.6\times 10^{51} $ & $5.9\times10^9$ & $4.78\times10^{-18}$   \\
Run-BP-M7   & broken power-law & $n_i=2.5, n_o=0$ & Figs \ref{fig1} and \ref{fig2} & $0.074$ & $3.6\times 10^{51} $ & Fig \ref{fig2} & Fig \ref{fig2} \\

\hline\noalign{\smallskip}
\end{tabular}}

Note: the wind properties shown in Figs \ref{fig1} and \ref{fig2} are from model M7 of the TDE simulation in \cite{BU22}. *$n$, $n_i$ and $n_o$ are the power-law index of the CNM profile (Equations (1) and (2)).
\end{table}

\subsection{Shock Capture and CRe Injection}

In this work, we adopt a CRe injection method similar to that of \cite{2026ApJ..1006...70M}, but with a different shock capture approach based on the Rankine-Hugoniot (RH) jump conditions. Following \cite{2026ApJ..1006...70M}, we define $\epsilon_{\rm e}$ as the fraction of the energy flux dissipated at the shock that is converted into CRe energy in the downstream region. In the shock front frame, we have
\begin{equation}
    \label{eq3}
    \epsilon_{\rm e} = \frac{e_2 \mu_{\rm d}}{0.5 \rho_{\rm u} \mu_{\rm u}^3 (1 - C^{-2})},
\end{equation}
where $e_2$ is the CRe energy density in the downstream region, $\mu_{\rm d}$ is the velocity normal to the shock front in the downstream, $\rho_{\rm u}$ is the upstream density, and $\mu_{\rm u}$ and $C$ are the velocity normal to the shock front in the upstream and the compression ratio, respectively. The subscripts $\rm u$ and $\rm d$ denote the upstream and downstream regions of the shock, respectively.

To locate the downstream region of the shock where CRe injection is needed, we adopt the shock detection method used in \cite{Lovely1999} and \cite{2025A&A...704A.148X}. We assume that the normal direction of the shock coincides with the direction of the pressure gradient, and project the flow velocity onto this direction as $u = \mathbf{v} \cdot \frac{\nabla p}{|\nabla p|}$, where p is the thermal pressure.

For each grid cell with a density exceeding the initial CNM density and a negative velocity divergence ($\nabla\cdot\mathbf{v}<0$, indicating compression), we treat it as a candidate for the downstream region of a shock. Along the direction opposite to the pressure gradient ($-\hat{e}_p \equiv -\nabla p / |\nabla p|$), we identify the corresponding upstream grid cell. To reduce the impact of numerical instabilities and exclude very weak shocks, we pre-exclude those grid pairs for which the pressure jump is small, i.e., $(p_d - p_u)/p_u < 2$. According to the Rankine-Hugoniot jump conditions, the upstream Mach number is given by
\begin{equation}
    \label{eq4}
    \mathrm{M_u} = \sqrt{\frac{\gamma_1+1}{2\gamma_1}\frac{p_d}{p_u} + \frac{\gamma_1-1}{2\gamma_1}},
\end{equation}
where $\gamma_1 = 5/3$ is the adiabatic index of the gas.

In the shock front frame, the upstream gas flows into the shock with a velocity of $\mu_{\rm u}=\mathrm{M_u} a_{\rm u}$, where $a_{\rm u} \equiv \sqrt{\gamma_1 p_{\rm u}/\rho_{\rm u}}$ is the upstream sound speed. In the lab frame, the positive direction is locally defined by the pressure gradient direction $\hat{e}_p$; the shock speed is then obtained by subtracting the velocity of the upstream gas in the shock front frame ($\mu_{\rm u}$) from the upstream gas velocity in the lab frame. The shock speed in the lab frame, $\mathrm{W}$, is therefore computed as $\mathrm{W} = u_{\rm u} - \mu_{\rm u}$.

For a valid shock wave, the upstream and downstream properties should satisfy the Lax entropy condition, namely $|\mu_{\rm u}| > a_{\rm u}$ (upstream supersonic) and $|\mu_{\rm d}| < a_{\rm d}$ (downstream subsonic). In the shock front frame, both of the upstream and downstream velocity are along the direction of the pressure gradient, thus both of them are positive according to our definition of the positive direction of this work. Therefore the Lax condition becomes
\begin{equation}
    \label{eq5}
    u_{\rm d} - a_{\rm d} < {\rm W} < u_{\rm u} - a_{\rm u}.
\end{equation}
Grid cells that satisfy condition Eq.\ref{eq5} are considered to be part of the shock dissipation region. Since numerical simulations tend to smear sharp shocks into extended shock regions, we identify the most downstream grid cells whose immediate downstream neighbors no longer satisfy Eq.\ref{eq5} as the effective downstream location, and inject CRe into these cells.

For each downstream grid cell, we trace its effective upstream grid cell along the direction opposite to the pressure gradient ($-\hat{e}_p$); this effective upstream cell is defined as the one whose immediate upstream neighbors no longer satisfy Eq.~\ref{eq5}. Using the effective upstream and downstream quantities, we calculate the effective upstream Mach number $\mathrm{M_u}$ via Eq.~\ref{eq4}. Then, according to the RH jump conditions, the compression ratio $C$ is derived as
\begin{equation}
    \label{eq6}
    C \equiv \frac{\rho_{\rm d}}{\rho_{\rm u}} = \frac{1}{\frac{\gamma_1-1}{\gamma_1+1} + \frac{2}{(\gamma_1+1)\mathrm{M_u}^2}}.
\end{equation}
Substituting the compression ratio $C$ and $\mu_{\rm u} = \mathrm{M_u} a_{\rm u}$ into Eq.~\ref{eq3}, and using the relation $\mu_{\rm u}/\mu_{\rm d} = C$, we inject CRe into these grid cells as
\begin{equation}
    \label{eq7}
    e_2 = \epsilon_{\rm e} \cdot \frac{1}{2} \rho_{\rm u} (\mathrm{M_u} a_{\rm u})^2 \left( C - \frac{1}{C} \right).
\end{equation}

The value of $\epsilon_{\rm e}$ is highly uncertain. While the cosmic ray proton (CRp) acceleration efficiency is $\sim 10\%$ for high-Mach-number shocks (e.g., \cite{2014ApJ...783...91C}), the CRe acceleration efficiency is expected to be lower. In this work, we adopt the same value as that in \cite{2026ApJ..1006...70M}, setting $\epsilon_{\rm e} = 0.03$.

To maintain energy conservation, when we inject CRe into the downstream region, we subtract an equivalent amount of gas thermal energy from these grid cells. Incorporating the CRe energy density, the final governing equations become
\begin{equation}
\begin{split}
\frac{\partial \rho}{\partial t} + \nabla \cdot (\rho \mathbf{v}) &= 0, \\
\rho \frac{d\mathbf{v}}{dt} &= -\nabla (p_1 + p_2) - \rho \nabla \Phi, \\
\frac{\partial e_1}{\partial t} + \nabla \cdot (e_1 \mathbf{v}) &= -p_1 \nabla \cdot \mathbf{v} - Sc, \\
\frac{\partial e_2}{\partial t} + \nabla \cdot (e_2 \mathbf{v}) &= -p_2 \nabla \cdot \mathbf{v} + Sc.
\end{split}
\end{equation}
Here, $e_1$ is the thermal energy density, $p_1 \equiv (\gamma_1-1)e_1$ is the thermal pressure, $p_2 \equiv (\gamma_2-1)e_2$ is the CRe pressure with adiabatic index $\gamma_2 = 4/3$, $\Phi = -GM_{\rm SMBH}/r$ is the gravitational potential, and $Sc$ represents the source term from the shock capture and CRe injection model.

\subsection{Radio Emission Calculation}

Given the simulated CRe distribution, we compute its synchrotron radiation using a method similar to that described in \cite{Rybicki1979}.

The post-processing domain is a cube in the Cartesian frame. To better resolve the expansion of the wind, we employ different calculation box size at different time durations. The length of the three axis are same, which is set to 0.033 pc at $t\le54.7\rm\ days$. The calculation box expands with the wind and the axis length increase to 0.04 pc at $54.7\ \mathrm{day}\le t\le87.6\rm\ day$; 0.056 pc at $87.6\ \mathrm{day}\le t\le120.4\rm\ day$; 0.74 pc at $120.4\ \mathrm{day}\le t\le147.8\rm\ day$; 0.135 pc at $147.8\ \mathrm{day}\le t\le300.0\rm\ day$; 0.39 pc at $300.0\ \mathrm{day}\le t\le1000.0\rm\ day$; 2.4 pc at $1000.0 \ \mathrm{day}\le t\le15\rm\ yr$. 

The line of sight is along the Z direction, which forms an angle $\theta_{\rm obs}$ with the rotational axis ($\theta = 0$) of our simulation frame. We calculate the final emission properties at the observer plane $Z = Z_{\rm image} = 2.4$~pc.

To convergence our calculation results, we employ 7432 uniformly spaced grids along the Z direction. In the X-Y plane, we employ $1190\times1190$ uniform grids at $t\le300\ \rm day$ and employ $768\times768$ uniform grids at $t\ge300\ \rm day$. At the observer plane, these grid cells in the X--Y plane define lines of sight (LoS). For each LoS, we perform a full radiative transfer calculation along the Z direction using the following method:

Firstly, we account for the time delay due to the finite speed of light. Each X--Y plane, located at a specific Z coordinate $Z(i)$, defines a photon crossing time $(Z_{\rm image}- Z(i))/c$, where $c$ is the speed of light. Thus, at an observer time $t_{\rm obs}$, the contribution from a grid cell with coordinate $Z(i)$ is evaluated using quantities at time $t_{\rm obs} -(Z_{\rm image}- Z(i))/c$ when performing the radiative transfer calculation.

Secondly, we map the grid point $(X', Y', Z', t_{\rm obs} - (Z_{\rm image}- Z')/c)$ in the calculation frame to the simulation grid $(r, \theta, t_{\rm obs} - (Z_{\rm image}- Z')/c)$. For a given observer angle $\theta_{\rm obs}$, we apply a rotation matrix between the Cartesian frames:
\begin{equation}
\begin{pmatrix} X \\ Y \\ Z \end{pmatrix}
=
\begin{pmatrix}
\cos\theta_{\rm obs} & 0 & \sin\theta_{\rm obs} \\[2pt]
0                    & 1 & 0                    \\[2pt]
-\sin\theta_{\rm obs} & 0 & \cos\theta_{\rm obs}
\end{pmatrix}
\begin{pmatrix} X' \\ Y' \\ Z' \end{pmatrix},
\end{equation}
where the Z direction is aligned with the $\theta = 0$ axis of the simulation frame. In this Cartesian frame, we convert the coordinates to spherical coordinates:
\begin{equation}
\begin{aligned}
R_c &= \sqrt{X^{2} + Y^{2}} \\[2pt]
r   &= \sqrt{R_c^{2} + Z^{2}} = \sqrt{X^{2} + Y^{2} + Z^{2}} \\[2pt]
\theta'' &=
\begin{cases}
\arctan\left(\dfrac{R_c}{Z}\right),              & Z > 0 \\[8pt]
\dfrac{\pi}{2},                                   & Z = 0 \\[8pt]
\arctan\left(\dfrac{R_c}{Z}\right) + \pi,         & Z < 0
\end{cases}
\end{aligned}
\end{equation}
where the $\phi$ coordinate is not needed because our simulation is axisymmetric. Additionally, since we assume the wind is symmetric about the midplane, for grid points with $\theta'' > \pi/2$, the CRe properties are the same as those at $\theta = \pi - \theta''$. That is,
\begin{equation}
\theta =
\begin{cases}
\theta'',                & \theta'' \le \dfrac{\pi}{2} \\[8pt]
\pi - \theta'',          & \theta'' > \dfrac{\pi}{2}.
\end{cases}
\end{equation}
Once we have the coordinates $(r, \theta)$ corresponding to the grid point $(X', Y', Z')$ in the calculation frame, we interpolate the quantities from the simulation grid at time $t_{\rm obs} - (Z_{\rm image}- Z')/c$ and use them to compute the radiative transfer.

Using these simulation quantities, we can then compute the synchrotron emission and perform the radiative transfer calculation. For a grid cell $(X, Y, Z)$ in the calculation domain with CRe energy density $e_2$, we assume a power-law electron distribution:
\begin{equation}
N(\gamma)\,d\gamma = K\,\gamma^{-p}\,d\gamma, \quad \gamma \in [\gamma_{\min}, \gamma_{\max}],
\end{equation}
where $\gamma$ is the Lorentz factor and $p$ is the power-law index, which we adopt as $p = 2.5$ in this work. The CRe distribution is set with $\gamma_{\min} = 2.0$ and $\gamma_{\max} = 2000$, following \cite{2026ApJ..1006...70M}.

The CRe energy density $e_2$ can then be expressed as
\begin{equation}
e_2 = \int_{\gamma_{\min}}^{\gamma_{\max}} (\gamma m_e c^2)\,K\gamma^{-p}\,d\gamma = K\,m_e c^2\,I_p,
\end{equation}
where
\begin{equation}
I_p = \int_{\gamma_{\min}}^{\gamma_{\max}} \gamma^{1-p}\,d\gamma =
\begin{cases}
\dfrac{\gamma_{\max}^{2-p} - \gamma_{\min}^{2-p}}{2-p}, & p \neq 2, \\[10pt]
\ln\dfrac{\gamma_{\max}}{\gamma_{\min}}, & p = 2.
\end{cases}
\end{equation}
The normalization parameter $K$ can then be derived as
\begin{equation}
K = \frac{e_2}{m_e c^2\,I_p}.
\end{equation}

In this work, we simply assume that the magnetic energy density in the post-shock region is proportional to the gas thermal energy density, i.e.,
\begin{equation}
\frac{B^{2}}{8\pi} = \epsilon_B \, p_1,
\end{equation}
where the fraction of the energy density $\epsilon_B$ is set to 0.1.

Following \cite{Rybicki1979}, the synchrotron emission coefficient $j_\nu$ and absorption coefficient $\alpha_\nu$ can then be expressed as
\begin{equation}
j_\nu = C_1 \, K \, B^{\frac{p+1}{2}} \, \nu^{-\frac{p-1}{2}}
\quad [\rm erg\,s^{-1}\,cm^{-3}\,Hz^{-1}\,sr^{-1}],
\end{equation}
\begin{equation}
\alpha_\nu = C_2 \, K \, B^{\frac{p+2}{2}} \, \nu^{-\frac{p+4}{2}}
\quad [\rm cm^{-1}],
\end{equation}
where $C_1$ and $C_2$ are the synchrotron radiation constants. They are given by
\begin{equation}
C_1 = \frac{\sqrt{3}\,e^{3}}{4\pi\,m_e c^{2}\,(p+1)} \,
a(p) \, A_j \,
\left(\frac{3e}{2\pi\,m_e c}\right)^{\frac{p-1}{2}},
\end{equation}
\begin{equation}
a(p) = \Gamma\!\left(\frac{p}{4}+\frac{19}{12}\right) \Gamma\!\left(\frac{p}{4}-\frac{1}{12}\right),
\end{equation}
\begin{equation}
A_j = \frac{\sqrt{\pi}}{2}\,
\frac{\Gamma\!\left(\dfrac{p+5}{4}\right)}{\Gamma\!\left(\dfrac{p+7}{4}\right)},
\end{equation}
\begin{equation}
C_2 = \frac{\sqrt{3}\,e^{3}}{8\pi\,m_e} \,
b(p) \, A_a \,
\left(\frac{3e}{2\pi\,m_e^{3} c^{5}}\right)^{\frac{p}{2}}
\left(m_e c^{2}\right)^{p-1},
\end{equation}
\begin{equation}
b(p) = \Gamma\!\left(\frac{3p+2}{12}\right) \Gamma\!\left(\frac{3p+22}{12}\right),
\end{equation}
\begin{equation}
A_a = \frac{\sqrt{\pi}}{2}\,
\frac{\Gamma\!\left(\dfrac{p+6}{4}\right)}{\Gamma\!\left(\dfrac{p+8}{4}\right)}.
\end{equation}

Using $j_\nu$ and $\alpha_\nu$, we solve the radiative transfer equation
\begin{equation}
\frac{dI_\nu}{ds} = j_\nu - \alpha_\nu I_\nu
\end{equation}
along each LoS. Finally, we sum the contributions from each LoS as
\begin{equation}
P_\nu = \sum_{(X',Y')} I_\nu(X', Y') \, \Delta A
\quad [\rm erg\,s^{-1}\,Hz^{-1}\,sr^{-1}],
\end{equation}
where $\Delta A$ is the area of each grid cell in the X--Y plane.

\section{RESULTS}

\begin{figure}
    \centering
    \includegraphics[width=1.0\linewidth]{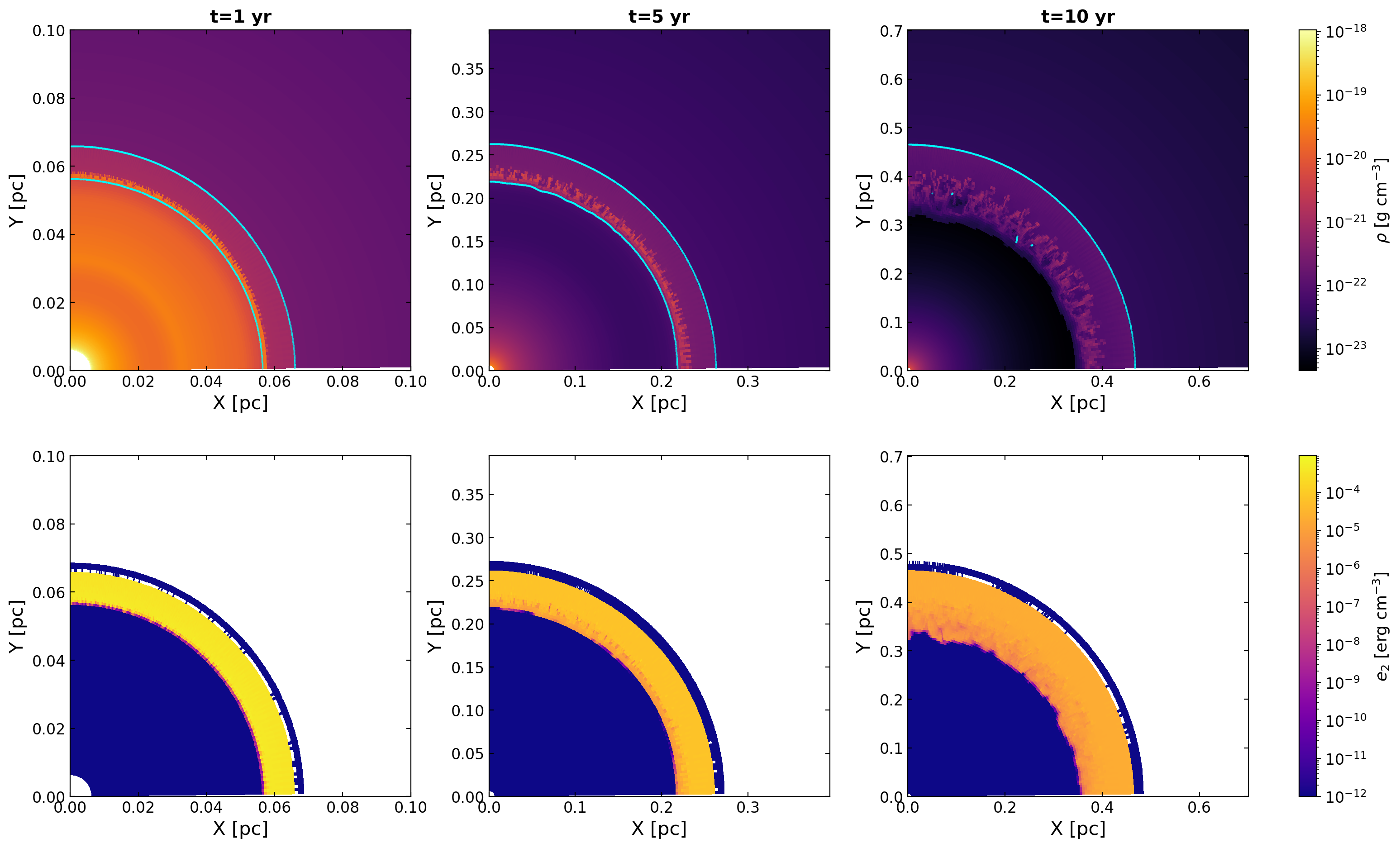}
    \caption{Snapshots of the gas density (top row) and CRe energy distribution (bottom row) from the simulation with spherical wind injection and a single power-law CNM profile (model Run-SP-sph). The left, middle, and right columns correspond to $t = 1$, $5$, and $10$~yr, respectively. In the top row, the blue lines indicate the downstream grid cells where we inject CRe, as identified by our shock capture model.}
    \label{fig3}
\end{figure}

Firstly, we present the gas and CRe energy density distributions and the associated hydrodynamic structures from our simulations. We adopt the spherically injected wind model with a single power-law CNM profile (model Run-SP-sph), as a representative of the uniform injection models. Fig.~\ref{fig3} shows snapshots of the gas density and CRe energy density at $t = 1$, $5$, and $10$~yr. In the density contours, the blue lines mark the downstream grid cells identified by our shock capture module (Eq.~\ref{eq5}), where CRe is injected.

At early times ($t = 1$~yr), our module identifies two distinct shock waves. One is a forward shock, whose upstream lies in the background CNM, while the other is a reverse shock that propagates through the dense wind. The injected CRe is distributed between these two shocks. At this stage, both shocks are strong, and their compression ratios reach the maximum value $C = 4$. In this regime, Eq.~\ref{eq3} can be simplified to $\epsilon_{\rm e} = 0.6\,e_2/e_1$ (\cite{2026ApJ..1006...70M}). With a constant $\epsilon_{\rm e} = 0.03$ adopted in this work, the CRe energy density is proportional to the thermal gas energy density in the downstream region. Since the thermal energy density is nearly uniform between the forward and reverse shocks, the CRe energy density is also uniformly distributed at early times (the left panel in the bottom row in Fig.~\ref{fig3}).

While the forward shock propagates through the CNM, the reverse shock travels through the dense and powerful wind. At late times (the right panel in the top row of Fig.~\ref{fig3}), our shock capture module identifies only a single forward shock. The most of the wind was injected within the first 200 days, so the bulk of the wind is confined within a finite radial width. Although the wind expands as it propagates to larger radii, at late times the reverse shock has already escaped the inner boundary of the wind and thus becomes weak and disappears. Consequently, CRe is not injected into the inner region at late times, and the CRe energy density $e_2$ in the inner region is slightly lower than that in the outer region, as shown in the right panel of the bottom row in Fig.~\ref{fig3}.

At $t = 1$~yr and $t = 5$~yr, a high-density structure appears near the reverse shock. The same feature is also seen in the simulations of \cite{2026ApJ..1006...70M} and  \cite{2026ApJ..1006..103L}. At late times, this structure becomes diffuse and turbulent. This morphological evolution is also observed in the simulations of \cite{2026ApJ..1006..103L}, where it is attributed to the Rayleigh--Taylor (RT) instability.

\begin{figure}
    \centering
    \includegraphics[width=1.0\linewidth]{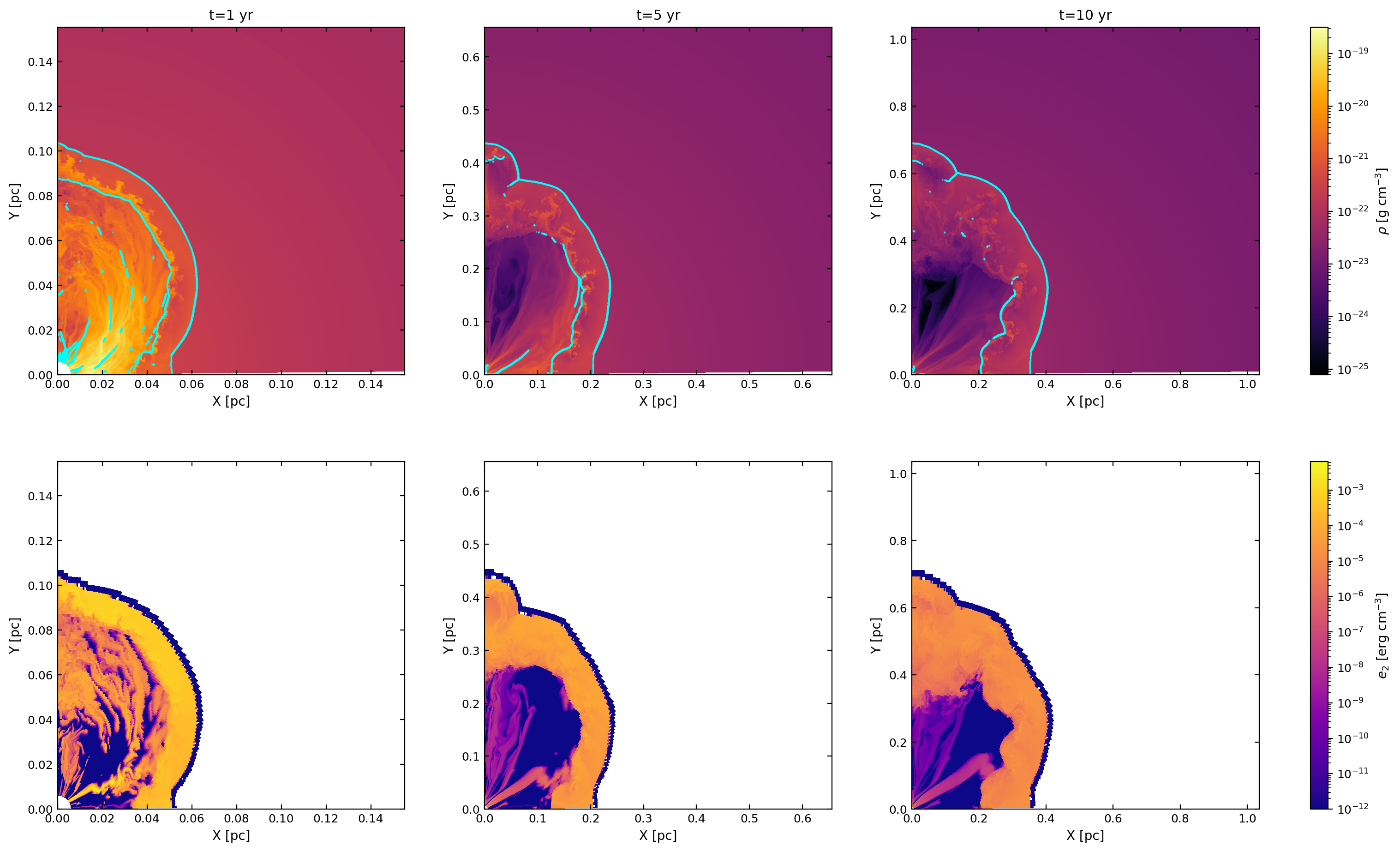}
    \caption{Same as Fig ~\ref{fig3}, but for model Run-SP-M7.}
    \label{fig4}
\end{figure}

The hydrodynamic behavior of all simulations with uniform, time-independent wind injection is the same as that described above. The morphology of the cone-injection simulations is slightly different from that of the spherical case, and our cone-injection results are broadly consistent with those of \cite{2026ApJ..1006..103L}. 

Now, we present the gas and CRe energy density structures from the simulation of model Run-SP-M7 in Fig.~\ref{fig4}.
As shown in Fig.~\ref{fig4}, the morphology of the expanding wind initially extends along the rotational axis (see the panels in the first and second columns of Fig.~\ref{fig4}) and then becomes slightly elongated near the mid-plane at late times (see the panels in the third column of Fig.~\ref{fig4}). The elongation along the rotational axis arises because the radial velocity of the injected wind is higher along the axis (see the top-right panel of Fig.~\ref{fig2}), while the slight morphological change at late times is due to the angle-dependent deceleration of the wind, which results from the non-uniform mass injection with angle (see the bottom-left panel of Fig.~\ref{fig2}). The second effect significantly impacts the radio light curves, as we will describe in the following subsection.

The simulation also exhibits both a forward shock and a reverse shock. While the reverse shock in model Run-SP-sph disappears at late times, the reverse shock in model Run-SP-M7 persists near the mid-plane, where the wind is denser and expands more slowly. Even at late times the reverse shock has not yet escaped the dense wind component near the mid-plane. Similar to model Run-SP-sph, most of the CRe is distributed between the two shocks.

As shown in Fig.~\ref{fig1}, the injected wind is quite time-variable, and we consequently observe numerous internal minor shocks in addition to the forward and reverse shocks (see the top-left panel of Fig.~\ref{fig4}). These internal minor shocks inject CRe into the system at early times (see the bottom-left panel of Fig.~\ref{fig4}). Since these internal shocks arise from the time-variable nature of the injected wind, their effect is significant at early times. At late times ($t > 200$ days), when we exponentially decay the injection of wind, these internal shocks begin to fade (see the middle and right panels in the top row of Fig.~\ref{fig4}). Without continuous injection, the CRe energy density in the inner region decreases rapidly with time due to adiabatic expansion. Consequently, at late times, the CRe energy density injected by the early internal shocks is lower than that injected by the outer main shocks by several orders of magnitude (see the middle and right panels in the bottom row of Fig.~\ref{fig4}). This indicates that the radio emission at late times is still dominated by the outer main forward shock, while the injection properties primarily affect the early-time radio emission. We will discuss this further in the following subsection.

\subsection{Radio Emission with a Single Power-law CNM Profile}

\begin{figure}
    \centering
    \includegraphics[width=1.0\linewidth]{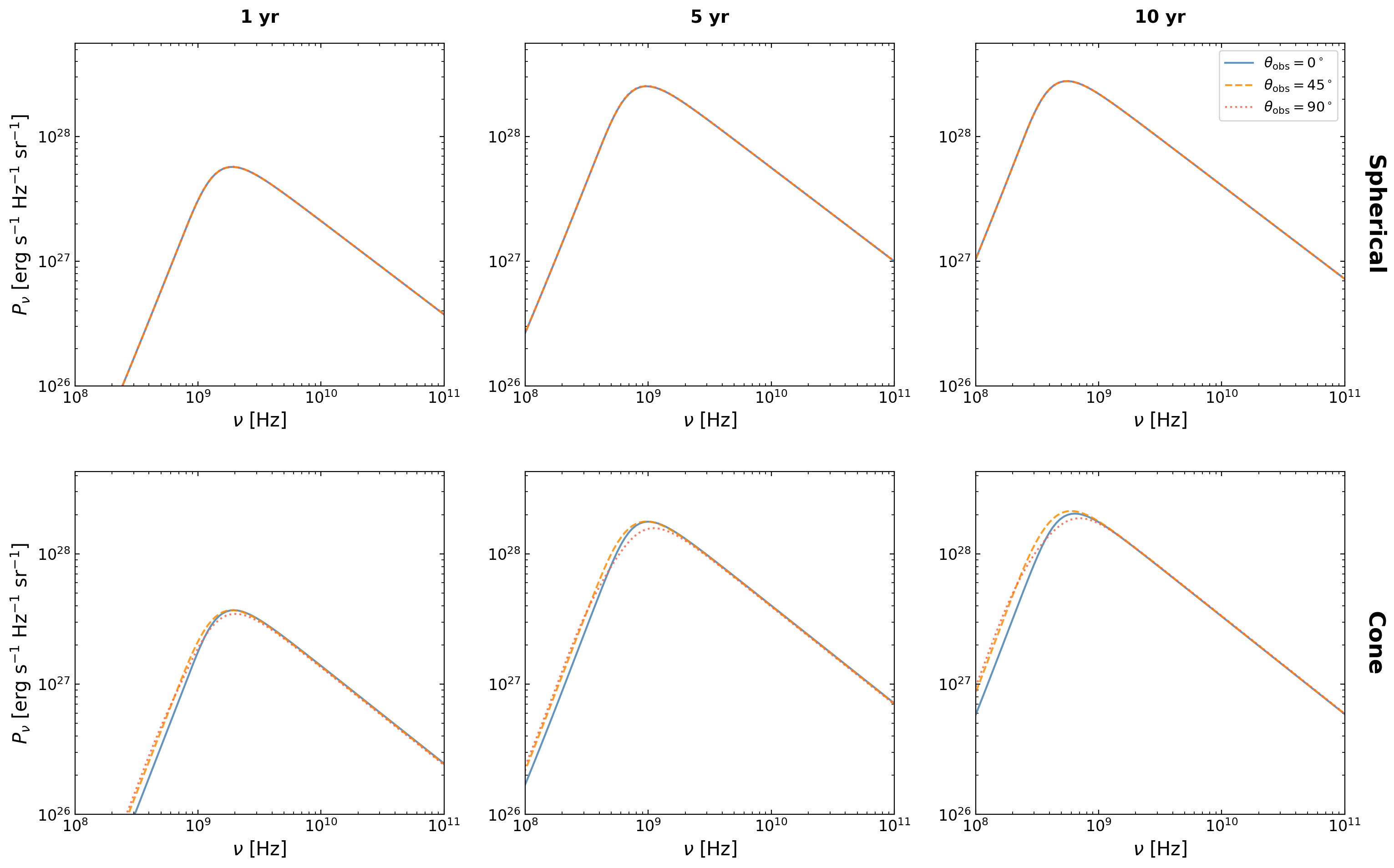}
    \caption{Temporal evolution of the radio SED for model Run-SP-sph (top row) and model Run-SP-cone (bottom row). The left, middle, and right columns correspond to $t = 1$, $5$, and $10$~yr, respectively. The blue solid, orange dashed, and red dotted lines indicate the SED along the LoS at $\theta_{\rm obs} = 0^\circ$, $45^\circ$, and $90^\circ$, respectively.}
    \label{fig5}
\end{figure}

\begin{figure}
    \centering
    \includegraphics[width=1.0\linewidth]{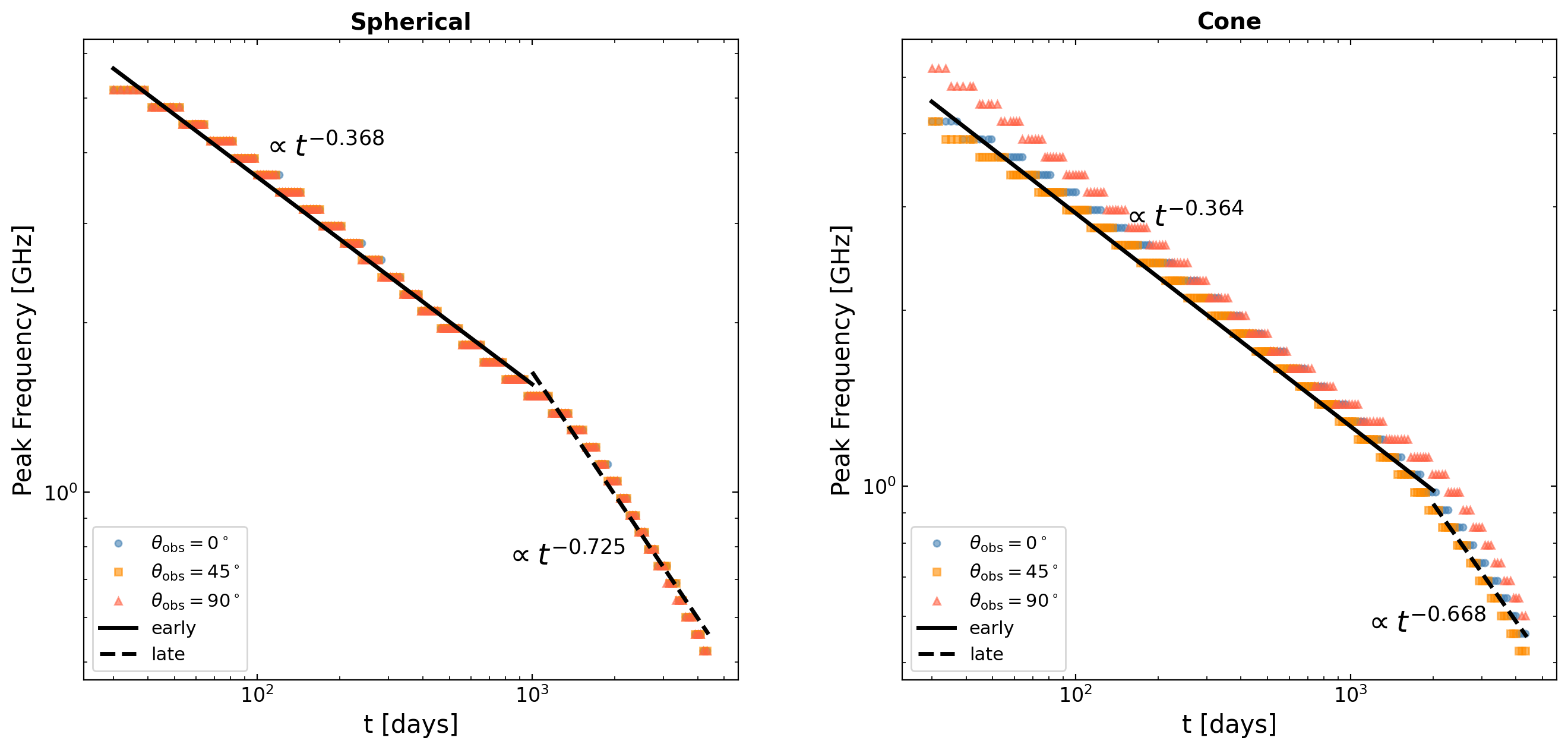}
    \caption{Temporal evolution of the peak frequency for model Run-SP-sph (left panel) and model Run-SP-cone (right panel). In each panel, the blue circles, orange squares, and red triangles mark the peak frequency at different times along the LoS at $\theta_{\rm obs} = 0^\circ$, $45^\circ$, and $90^\circ$, respectively. The solid and dashed lines in each panel show the power-law fits to the early and late phases, respectively. }
    \label{fig6}
\end{figure}

We first present the results from the simulations adopting the single power-law CNM profile described by Eq.~\ref{eq1}. Figure~\ref{fig5} shows the radio SEDs of the two uniform injection models (model Run-SP-sph and Run-SP-cone) at different times. The radio SED exhibits a characteristic double power-law shape, with the peak frequency $\nu_{\rm p}$ decreasing from GHz to sub-GHz scales over time (see Fig.~\ref{fig6}). The peak emission power can reach $10^{27-28}~{\rm erg\,s^{-1}\,Hz^{-1}\,sr^{-1}}$. The peak emission power from the model Run-SP-sph is slightly higher than that from the model Run-SP-cone. We attribute this difference to the distinct wind opening angle. The wind in Run-SP-sph has larger opening angle and can sweep more CNM, which affects the CRe injection amount and consequently modifies the SED.

There is a slight angular dependence of the SED for the model Run-SP-cone. In the optically thick regime, the emission power along a LoS with a larger $\theta_{\rm obs}$ (e.g., $\theta_{\rm obs} = 90^\circ$) is slightly higher than that with a smaller $\theta_{\rm obs}$ (e.g., $\theta_{\rm obs} = 0^\circ$). This phenomenon is also seen in the simulations of \cite{2026ApJ..1006...70M}, and this anisotropy becomes increasingly significant as the outflow opening angle decreases. As explained in \cite{2026ApJ..1006...70M}, in model Run-SP-cone, CRes are mainly present at the `cap' of the forward shock (the region within the wind's open angle). The lateral forward shock also host substantial CRe (see Fig. 3 in \cite{2026ApJ..1006...70M}). The anisotropic spectra is due to the anisotropic distribution of CRes.  Along the larger $\theta_{\rm obs}$, CRes in the lateral region contribute to lower frequency emission, leading to the broadening of the spectra below the peak frequency.

Fig.~\ref{fig6} provides further information on the temporal evolution of the peak frequency $\nu_{\rm p}$. The evolution of $\nu_{\rm p}$ exhibits two distinct phases. This is similar to the findings of \cite{2026ApJ..1006...70M}. The change in the evolutionary behavior is due to the significant deceleration of the wind at late times. At early times, when the wind deceleration is weak, the theoretical power-law index for the temporal evolution of $\nu_{\rm p}$ is $\sim -0.346$ for a CNM with a power-law index of $-1$, following the analytical work \cite{2021MNRAS.507.4196M}. In our results, we find indices of $-0.368$ for the model Run-SP-sph and $-0.364$ for the model Run-SP-cone, which decay slightly faster than the theoretical prediction. At late times, $\nu_{\rm p}$ decays with power-law indices of $-0.725$ and $-0.668$ for the model Run-SP-sph and cone injection models, respectively.

\begin{figure}
    \centering
    \includegraphics[width=1.0\linewidth]{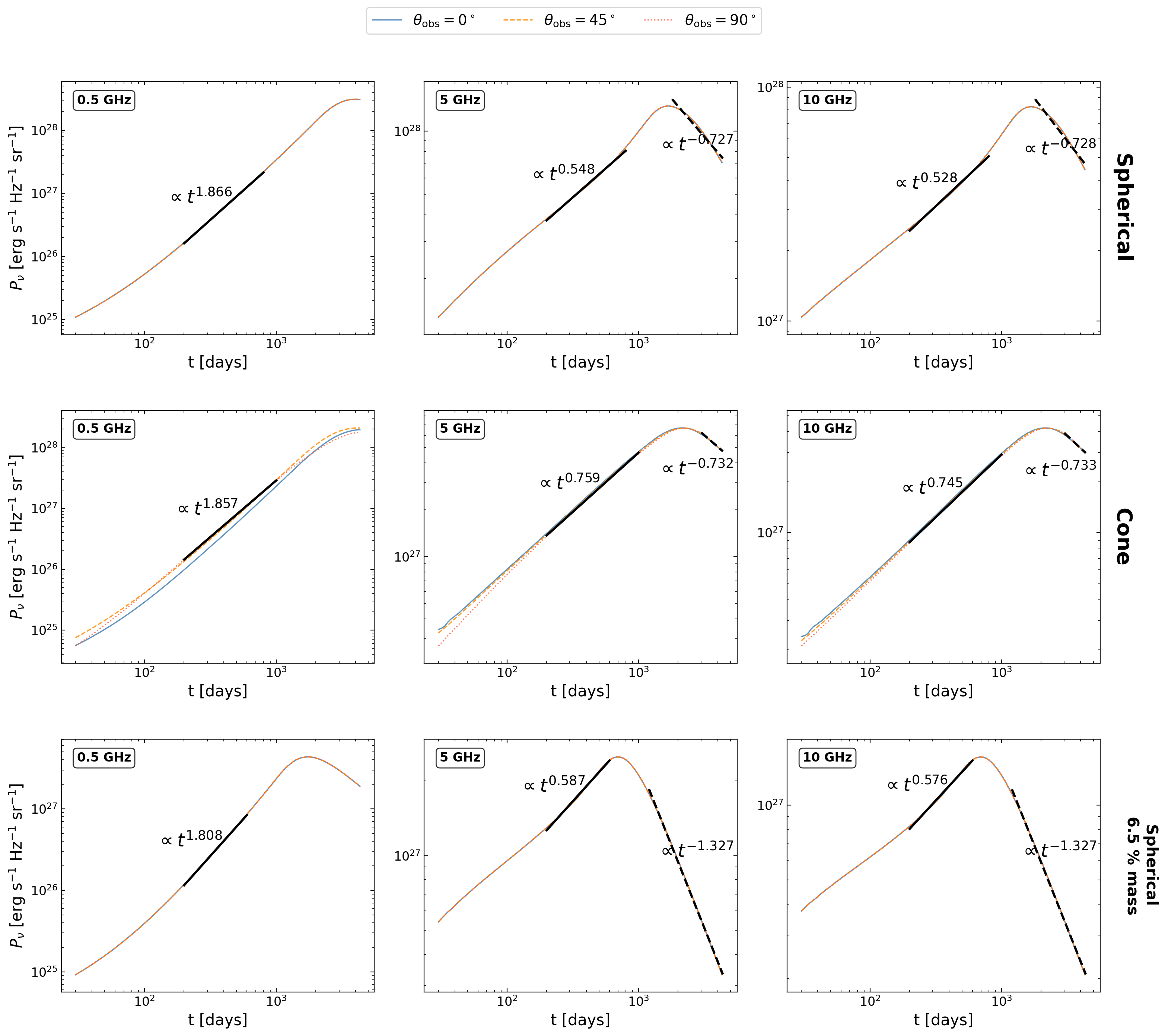}
    \caption{Light curves at 0.5 GHz (left), 5 GHz (middle), and 10 GHz (right) for model Run-SP-sph (top row), model Run-SP-cone (middle row), and model Run-SP-sphL that injects 6.5\% of the mass of model Run-SP-sph (bottom row). In each panel, the blue solid, orange dashed, and red dotted lines indicate the light curves along the LoS at $\theta_{\rm obs} = 0^\circ$, $45^\circ$, and $90^\circ$, respectively. The black solid and dashed lines show the power-law fits to the early and late phases, respectively.}
    \label{fig7}
\end{figure}

In addition to the SED, Fig.~\ref{fig7} presents the light curves at different frequencies for the two uniform injection models and a test model Run-SP-sphL that spherically injects 6.5\% of the mass of Run-SP-sph. The light curve at each frequency exhibits an initial rise followed by a late-time decline. The light curve at 5 GHz is similar to that at 10 GHz, while the 0.5 GHz light curve differs from both of them.

We first discuss the late-time decline. This decline is caused by the deceleration of the wind at late times. The onset of the light-curve decline roughly corresponds to the time when the mass of CNM swept up by the wind becomes comparable to the wind mass. For the test model Run-SP-sphL, in which we inject a lower mass than in our main models, this occurs at $700$~days at both 5 and 10 GHz. After a transition phase, the light curve decays as a power law, with indices of $-1.327$ at both 5 and 10 GHz. \cite{2026ApJ..1006...70M} derived the temporal evolution at late times analytically, giving the time index during the deceleration stage as $-\frac{3p+3}{2(5-n)}$, where $p$ is the power-law index of the CRe distribution and $n$ is the power-law index of the CNM profile. Substituting $p = 2.5$ and $n = 1$ yields a predicted index of $-1.3125$. Our simulations show a slightly steeper decline than the analytical prediction, and \cite{2026ApJ..1006...70M} reports the same faster decay. For the 0.5 GHz band, however, the decline begins later, at $t \sim 1700$~days. This is because the emission at 0.5 GHz is optically thick at early times, and the expansion of the photosphere strongly affects the overall emission. Even after the wind begins to decelerate significantly, the 0.5 GHz light curve continues to rise until the emission at this frequency becomes optically thin.

For models Run-SP-sph and Run-SP-cone, the injected mass is larger, which makes the decline of the light curves start later. As a result, for the optically thin 5 GHz and 10 GHz bands, the light curve begins to decay at $t \sim 1658$~days for model Run-SP-sph and at $t \sim 2265$~days for model Run-SP-cone. The later decline in model Run-SP-cone is because the solid angle of the cone injection is significantly smaller, so the wind needs to expand to larger radii to sweep up enough CNM to decelerate. However, the decay indices are $\sim -0.7$ at optically thin frequencies, which are significantly lower than those for the test model. This is because the deceleration timescale is close to the end of the simulation. Since the deceleration is weak, the wind undergoes a prolonged transition phase, during which the decline is gradual and becomes faster over time. Within our 15-yr simulation time, we are unable to trace the light curve until it reaches the stable self-similar regime described by the analytical expression.

For the initial rise, \cite{2021MNRAS.507.4196M} provides a comprehensive analytical treatment assuming that the wind velocity remains constant during the early phase. In the optically thin regime, for a CNM profile with power-law index $-1$ and a CRe distribution with power-law index $p = 2.5$, the analytical expression predicts a light-curve rise with a power-law index of $1.125$. In our simulations, at optically thin frequencies, the rise indices are $0.548$ at 5 GHz and $0.528$ at 10 GHz for the model Run-SP-sph, and $0.759$ at 5 GHz and $0.745$ at 10 GHz for the model Run-SP-cone. In the optically thick regime (0.5GHz), our two models yield rise indices of $1.866$ and $1.857$, respectively, while the analytical work \cite{2021MNRAS.507.4196M} gives $2.25$.

\begin{figure}
    \centering
    \includegraphics[width=0.8\linewidth]{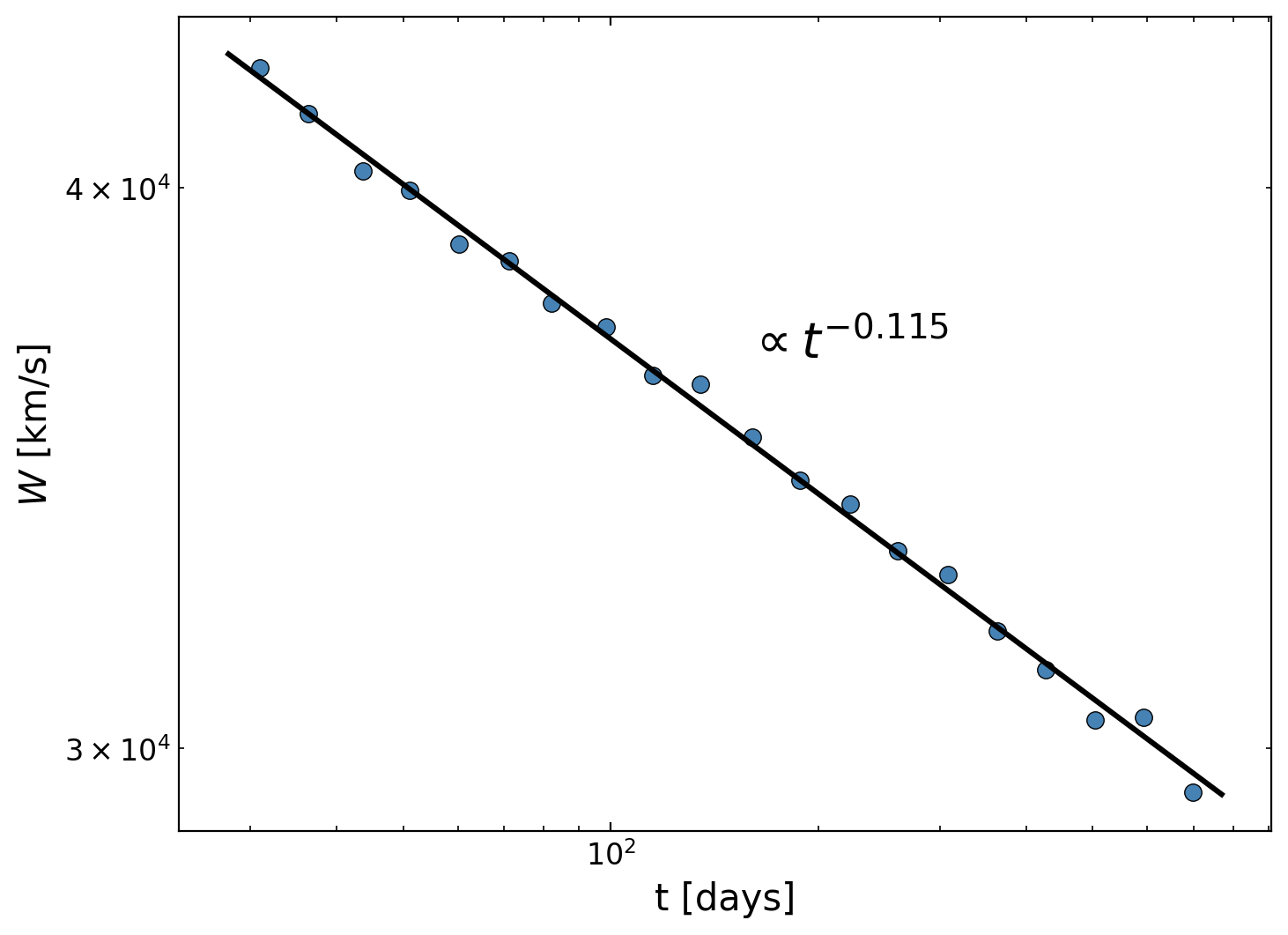}
    \caption{Evolution of the forward shock velocity $W$ in the model Run-SP-sph. The black line shows the power-law fit.}
    \label{fig8}
\end{figure}

The simulations of \cite{2026ApJ..1006..103L} and \cite{2026ApJ..1006...70M} also report power-law indices lower than unity in the optically thin regime for a CNM profile with $n = 1$. While \cite{2026ApJ..1006...70M} attributes this to adiabatic expansion of the CRe, \cite{2026ApJ..1006..103L} explains the phenomenon as resulting from ``the broader spatial expansion of the outflow and the non-negligible kinetic energy dissipation that occurs before the outflow reaches the nominal deceleration radius.'' In our simulations, we also find a slight decrease in the shock velocity calculated by our shock capture module at early times. Fig.~\ref{fig8} shows the forward shock velocity $\rm W$ at different times before 700 days. Despite minor fluctuations, the shock velocity decays roughly as $\sim t^{-0.115}$.

Incorporating this temporal evolution of the shock speed into the analytical framework of \cite{2021MNRAS.507.4196M}, we denote the CNM profile as $\rho \propto r^{-b}$ and the deceleration as $W \propto t^{s}$, which changes the radius evolution to $r \propto t^{\delta}$, where $\delta = 1 + s$. The change in the radius evolution would change the swept-up CNM number and the magnetic field evolution to $N_e \sim n r^3 \propto t^{\delta(3-b)}$ and $B \propto \sqrt{p_1} \propto \sqrt{n W^2} \propto t^{s-(b/2)\delta}$. Finally, in the deep Newtonian phase, the optically thin light curve follows $F_\nu \propto t^{\delta(3-b) + 2s + ((p+1)/2)(s-(b/2)\delta)}$. Substituting $s = -0.115$, $b = 1$, and $p = 2.5$, this expression yields a rising index of $\propto t^{0.564}$ in the optically thin regime, which is close to our model run-SP-sph. The analytical method developed by \cite{2021MNRAS.507.4196M} describes the effect of radial expansion; thus, our spherical model is consistent with it. However, under similar deceleration, the rise of the optically thin light curve of model run-SP-cone is faster. We attribute this to the expansion in the $\theta$ direction, which injects more CRe than the analytical prediction, since we find that calculating the light curve using only grid cells with $\theta < 30^\circ$ yields a light curve similar to that of run-SP-sph.

A simple assumption of momentum conservation during the early sweeping of the CNM would give an extremely slow deceleration, such that the velocity could be regarded as roughly constant. However, in our simulations, the wind is injected over a finite time interval, which differs from the impulsive injection scenario; the forward shock is thus driven only by a fraction of the total kinetic energy. Recent simulations by \cite{2026arXiv260703548D} also report that the velocity of a continuously injected outflow decreases and deviates from the ideal Blandford--McKee self-similar solution at early `free expansion phase'.

Additionally, continuous injection may accelerate the wind, as the decelerated wind is pushed by wind injected at later times. The simulation of \cite{2025ApJ...988L..24H} finds that its wind experiences an acceleration phase. In our simulations, we do not find such acceleration, but we find that the early deceleration of the wind becomes slightly slower at late times (e.g., $s$ becomes larger). The optically thin light curve shown in Fig.~\ref{fig7} exhibits a faster rise during $\sim 700~\mathrm{days} < t < 1658~\mathrm{days}$, with the power-law index increasing to $\sim 0.7$. This indicates that the deceleration becomes slower. This phenomenon is not significant in model run-SP-cone because the light curve is significantly affected by the $\theta$ expansion, which overwhelms the slight slowdown of the deceleration effect.

Despite the different injection properties, the results from our uniform, time-independent wind injection models are broadly consistent with previous analytical and numerical studies. In contrast, the results from model Run-SP-M7 show notable differences.

\begin{figure}
    \centering
    \includegraphics[width=1.0\linewidth]{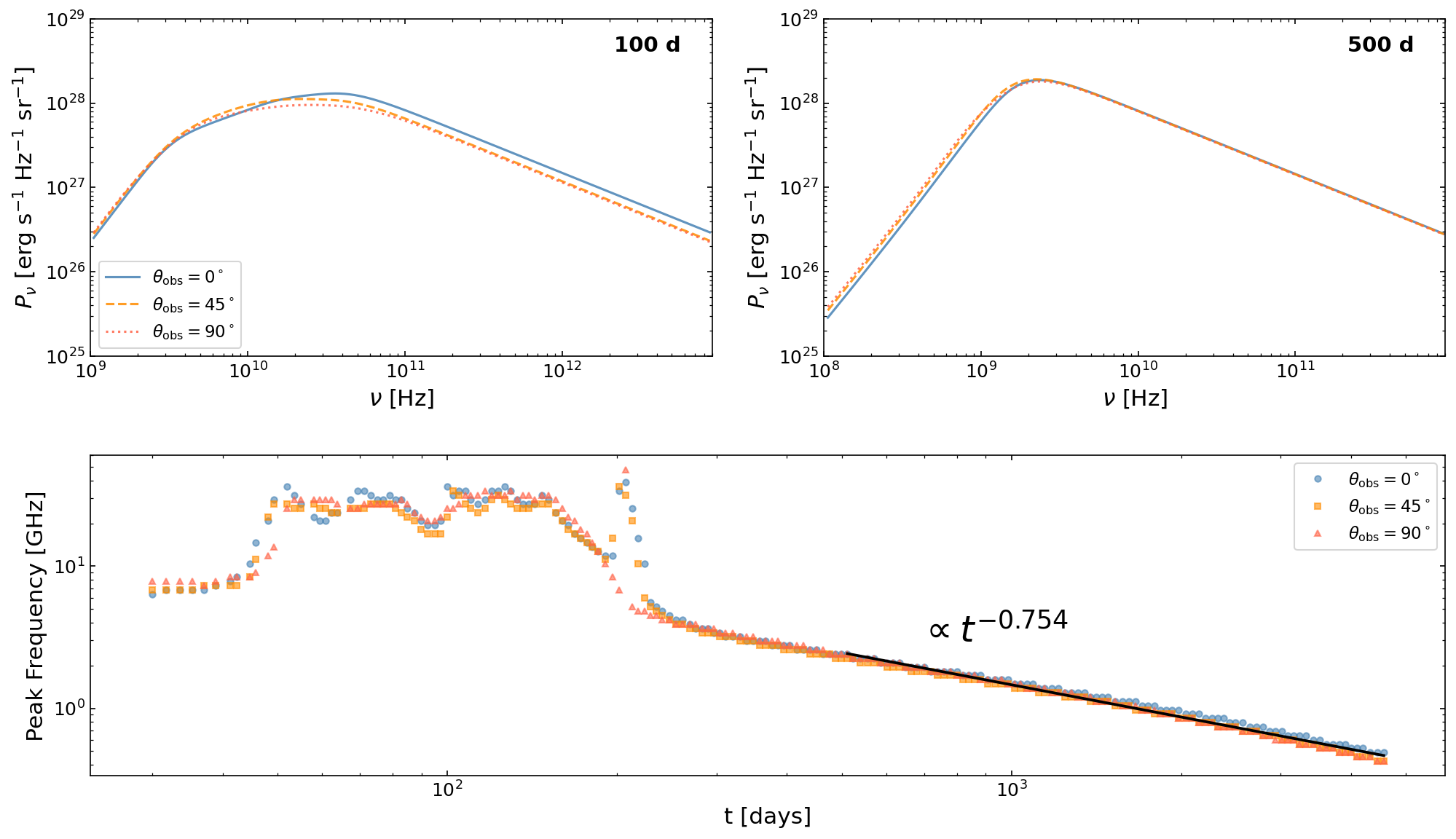}
    \caption{The top panels show the SED of the model Run-SP-M7 at $t = 100$~days (left) and $t = 500$~days (right). The bottom panel shows the temporal evolution of the peak frequency $\nu_{\rm p}$. The format is the same as in previous figures.}
    \label{fig9}
\end{figure}

The top panels of Fig.~\ref{fig9} show the SED of our model Run-SP-M7 at an early snapshot ($t = 100$~days) and a slightly later snapshot ($t = 500$~days); the bottom panel shows the temporal evolution of the peak frequency $\nu_{\rm p}$. While the uniform injection simulations exhibit a typical double power-law synchrotron SED, the SED of model Run-SP-M7 shows a double components spectrum at early times. The first component, which has a lower peak frequency and is similar to that in the uniform injection case, shows little variation across different LoSs, due to the outer main forward shock. The second component, however, varies with the LoS. This arises from the CRe injected by the inner shocks caused by the time variable nature of the injected wind. These shocks are unstable, and the spatial distribution of the CRe injected by them is highly spatial non-uniform and rapidly evolving. The path that a LoS traverses through these CRe clumps differs with the observer angle $\theta_{\rm obs}$, leading to a LoS-dependent optical depth. Consequently, both the location and strength of the second component vary with the LoS and evolve rapidly. The second component evolves rapidly around $t=\rm 200\ days$. This is affected by our unphysical discontinuous wind that we choose to cut off the wind injection at this time. This approach does not affect our main results of this work because its impact disappeared rapidly. At late times, when the injection decays exponentially, these inner shocks gradually disappear, and the CRe energy density injected by them decreases due to adiabatic expansion, as described above. The SED then settles into a characteristic single-peaked shape, dominated by the forward shock that remains strong. Thereafter, the temporal evolution of $\nu_{\rm p}$ becomes stable, and its power-law index -0.754.

These inner shocks, whose strength is affected by the time-variable nature of the injected wind, are partly affected by our simulation setup, which relies on the results of the TDE simulation in \cite{BU22}. The output time interval of the previous simulation is much larger than that required by the CFL condition in the present radio simulations. We time-interpolate the wind properties during injection, this approach smooths the time-variability of wind. However, we do not know to what extent the output wind time-variability deviates from the true time-variability of wind. We must note that actually the super-Eddington accretion flow is quite turbulent, the resulting wind in nature is time-variable. Although, quantitatively, the injected time-variable nature of wind may deviate from the real case due to the finite output time-interval of the simulations of \cite{BU22}, the qualitative nature of the wind time-variability is captured. Moreover, in these inner regions, due to the instability of the inner shocks, a fluid element may cross multiple shocks and be accelerated multiple times. The value of $\epsilon_{\rm e}$ may differ for electrons experiencing different acceleration episodes, and our current understanding of such multi-acceleration processes is limited. Therefore, the results from our simulations regarding the inner shocks have limited representativeness. 

The inner shocks only affect the system at early times when the wind is injected, while the long-term CRe properties are determined by the outer forward shock. Thus, the main focus of this work is on the effects of the simulated wind at longer timescales. Although the SED shows negligible differences at late times, the light curves exhibit more significant changes.

\begin{figure}
    \centering
    \includegraphics[width=1.0\linewidth]{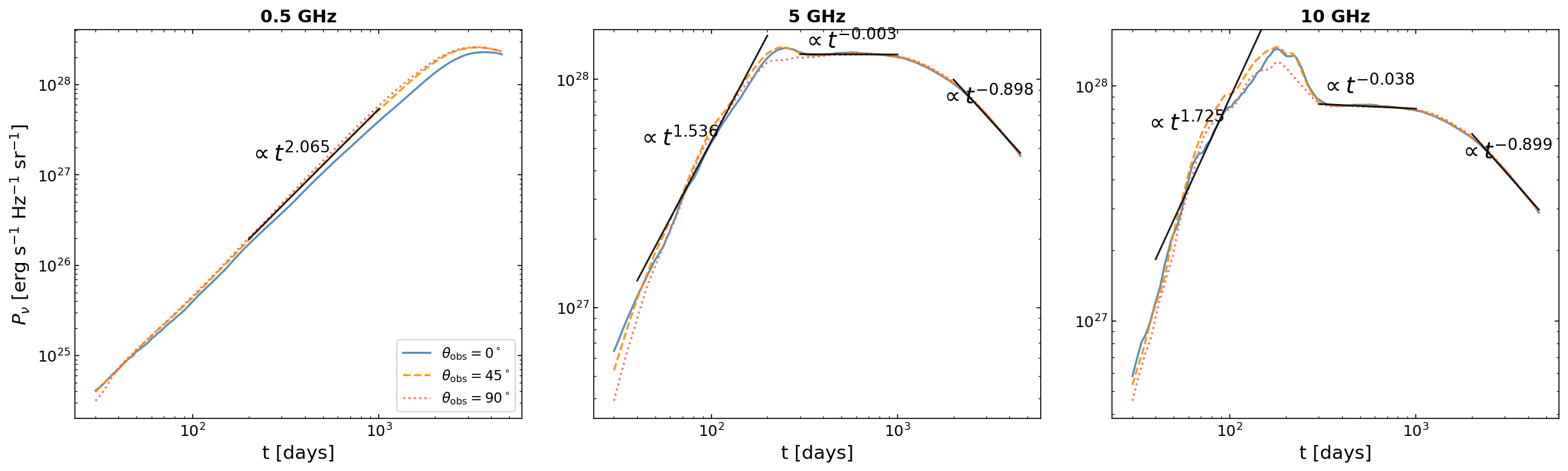}
    \caption{Light curves at 0.5 GHz, 5 GHz, and 10 GHz for the model Run-SP-M7.}
    \label{fig10}
\end{figure}

Fig.~\ref{fig10} shows the light curves at 0.5 GHz, 5 GHz, and 10 GHz for model Run-SP-M7. At 0.5 GHz, the emission remains optically thick and continues to rise until $t \sim 3000$~days, with a power-law index of 2.065, close to the theoretical prediction for the optically thick regime (\cite{2021MNRAS.507.4196M}). At higher frequencies, which are optically thin for the forward shock component but still optically thick for the inner shock component, the early rise is faster than in the uniform injection case (with indices of $\sim 1.536$ at 5 GHz and $\sim 1.725$ at 10 GHz). At $t \sim 200$~days, the peak frequency of the inner shock component crosses 10 GHz, and this component is turbulent, as shown in Fig.~\ref{fig9}; therefore, the light curve at 10 GHz is quite unstable around this time. After $t \sim 300$~days, the contribution from the inner shocks decreases significantly, making the light curve decrease significantly; after that, the light curve becomes dominated by the main forward shock. However, while the uniform injection models show a rise with a power-law index of $\sim 0.5-0.7$ (Fig.~\ref{fig7}) in the optically thin regime, model Run-SP-M7 exhibits a plateau phase between $\sim 300$ and $\sim 1000$~days (with indices of $-0.003$ at 5 GHz and $-0.038$ at 10 GHz during this period), followed by a slow decline. Eventually, it reaches a state with power-law indices of $\sim -0.9$. Similar to the uniform injection models, the late-time decline steepens with time; if we fit only the data between 3500 days and 15 yr, the index decreases to $-1.002$.

\begin{figure}
    \centering
    \includegraphics[width=0.8\linewidth]{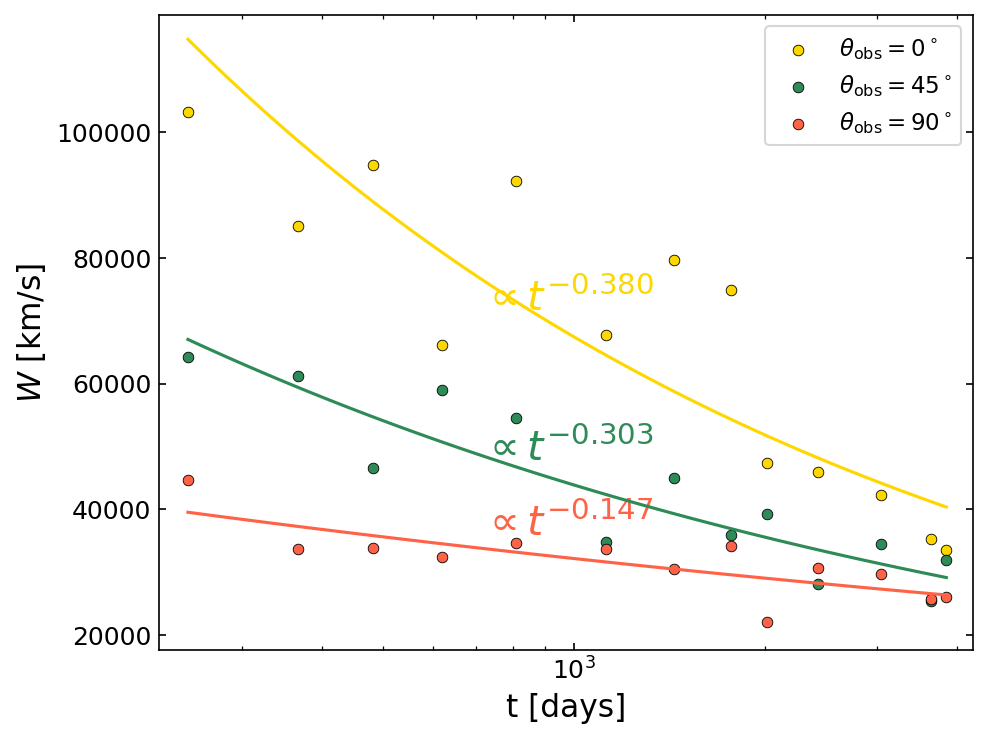}
    \caption{The temporal evolution of the forward shock velocity W in the model Run-SP-M7 at $\theta = 0^\circ$ (yellow) $\theta = 45^\circ$ (green) $\theta = 90^\circ$ (red) dots, and the solid lines denote the fitting assuming a simple power-law.} 
    \label{fig11}
\end{figure}

This discrepancy in the optically thin light curve arises from the different deceleration processes along different $\theta$ directions. Fig.~\ref{fig11} shows the temporal evolution of the forward shock speed along $\theta = 0^\circ$, $45^\circ$, and $90^\circ$. Despite some turbulence, and although a simple power-law fit cannot fully describe the deceleration over the entire time range, we can still see that the deceleration at $\theta = 0^\circ$ is faster than that at $\theta = 90^\circ$.

As discussed above, the deceleration of the wind affects the light curve. The slower the velocity decreases, the stronger the CRe injection is and the faster the light curve rises. Even with the same injected mass and kinetic energy, the spatial distribution alters the deceleration process, including both the time at which the wind begins to decelerate significantly and the magnitude of the deceleration itself. For model Run-SP-M7, the injection velocity and mass vary significantly with angle. Consequently, for an angle-independent CNM profile, the deceleration differs significantly across angles. Compared to a situation in which the wind decelerates slowly in all directions, this angle-dependent deceleration is steeper along the rotational axis. Therefore, the light curve would have a smaller power-law index if we consider only the gas along the axis, and a larger power-law index if we consider only the gas around the disk. Finally, integrating the contributions from all angles, the resulting light curve would differ from any single angular component.

\subsection{Light Curves with a Broken Power-law CNM Profile}

While the light curves at optically thick frequencies show only weak dependence on the CNM profile, those in the optically thin regime are significantly affected by the CNM distribution. As the power-law index of the CNM profile decreases, the rise of the optically thin light curve becomes slower, and may even turn into a decay if the CNM density declines sufficiently rapidly with radius. To show this more clearly, we run two additional simulations (Run-SP-cone-0 and Run-SP-cone-2.5) with single power-law CNM profile.

\begin{figure}
    \centering
    \includegraphics[width=1.0\linewidth]{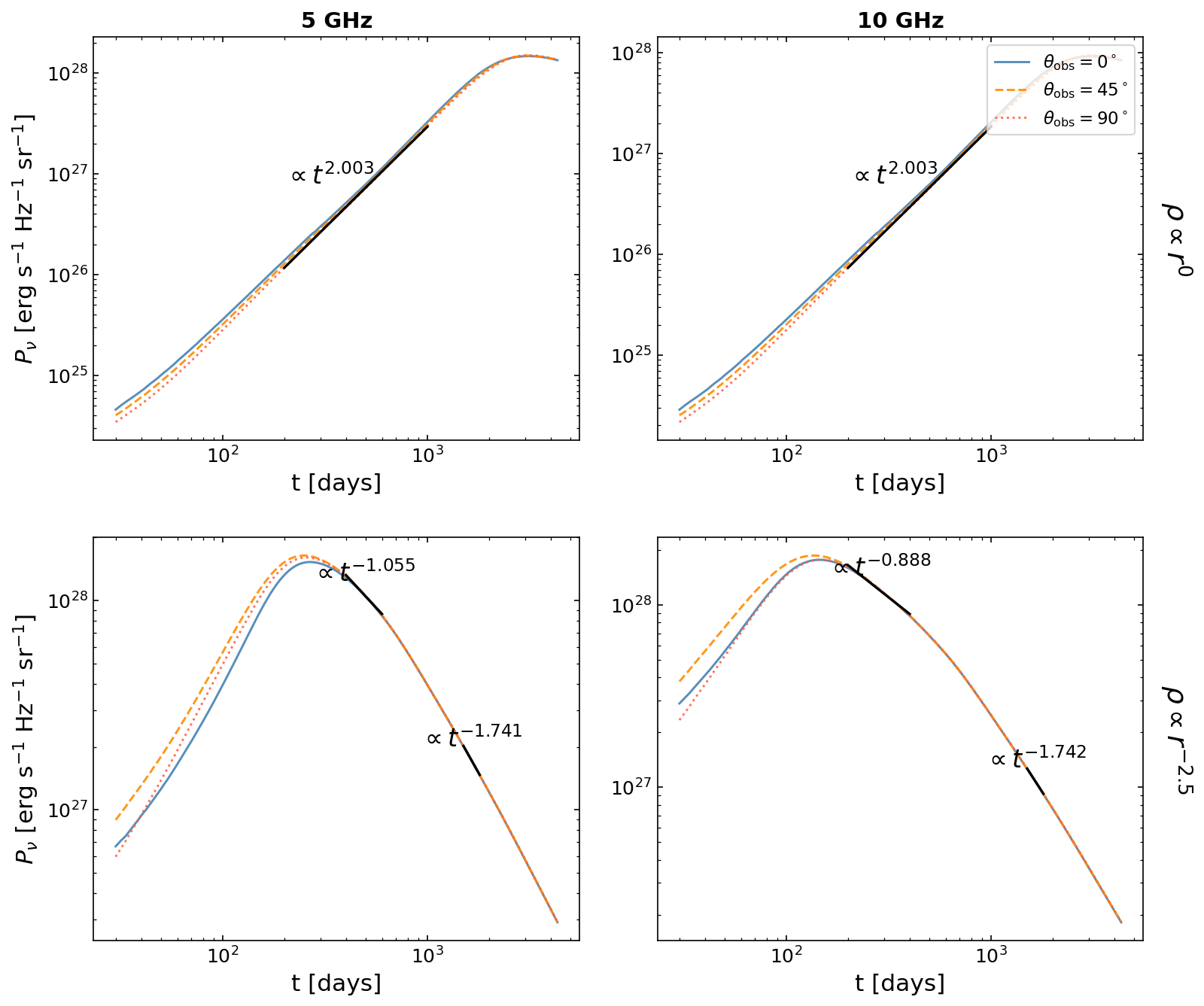}
    \caption{Light curves at 5 GHz (left column) and 10 GHz (right column) for the cone injection model with a single power-law CNM profile of index 0 (model Run-SP-cone-0, top row) and $-2.5$ (model Run-SP-cone-2.5, bottom row). The blue solid, orange dashed, and red dotted lines indicate the light curves along the LoS at $\theta_{\rm obs} = 0^\circ$, $45^\circ$, and $90^\circ$, respectively.}
    \label{fig12}
\end{figure}

 Fig.~\ref{fig12} shows the light curves for models Run-SP-cone-0 and Run-SP-Cone-2.5. Here we focus on the 5 GHz and 10 GHz bands, which remain in the optically thin regime for a longer duration. In the optically thin phase, before the wind has decelerated significantly, the light curve with $\rho \propto r^0$ rises with a power-law index of $\sim 2$. This rise is faster than that in the simulations presented in the previous subsection (see Fig.~\ref{fig7}). This is because the slower the CNM density decreases with radius, the more kinetic energy of the upstream gas is retained as the shock expands, assuming that the shock velocity does not change significantly; thus, the wind injects more CRes during this phase. This rise is slower than the analytical prediction (\cite{2021MNRAS.507.4196M}) but is consistent with the simulations of \cite{2026ApJ..1006..103L} for a $\rho \propto r^0$ CNM profile. As discussed above, we attribute this to the slow deceleration of the wind.

On the other hand, the light curve with $\rho \propto r^{-2.5}$ shows a decline after its frequency becomes optically thin. The onset time of this decline differs with frequency. As shown in Fig.~\ref{fig12}, the 5 GHz light curve begins to decay at $t \sim 300$~days, while the 10 GHz light curve starts to decay earlier, at $t \sim 130$~days, because it becomes optically thin sooner. After this, the light curve undergoes a transition phase, during which the power-law index is $\sim -0.9$. The decline then reaches a stable state at later times, with power-law indices of $-1.741$ and $-1.742$ at 5 GHz and 10 GHz, respectively. Since the CNM density following $\rho \propto r^{-2.5}$ is sufficiently low at large radii, the deceleration of the wind at late times is negligible. Consequently, the late-time power-law indices are consistent with both the analytical prediction (\cite{2021MNRAS.507.4196M}) and the simulation results of \cite{2026ApJ..1006..103L}.

\begin{figure}
    \centering
    \includegraphics[width=1.0\linewidth]{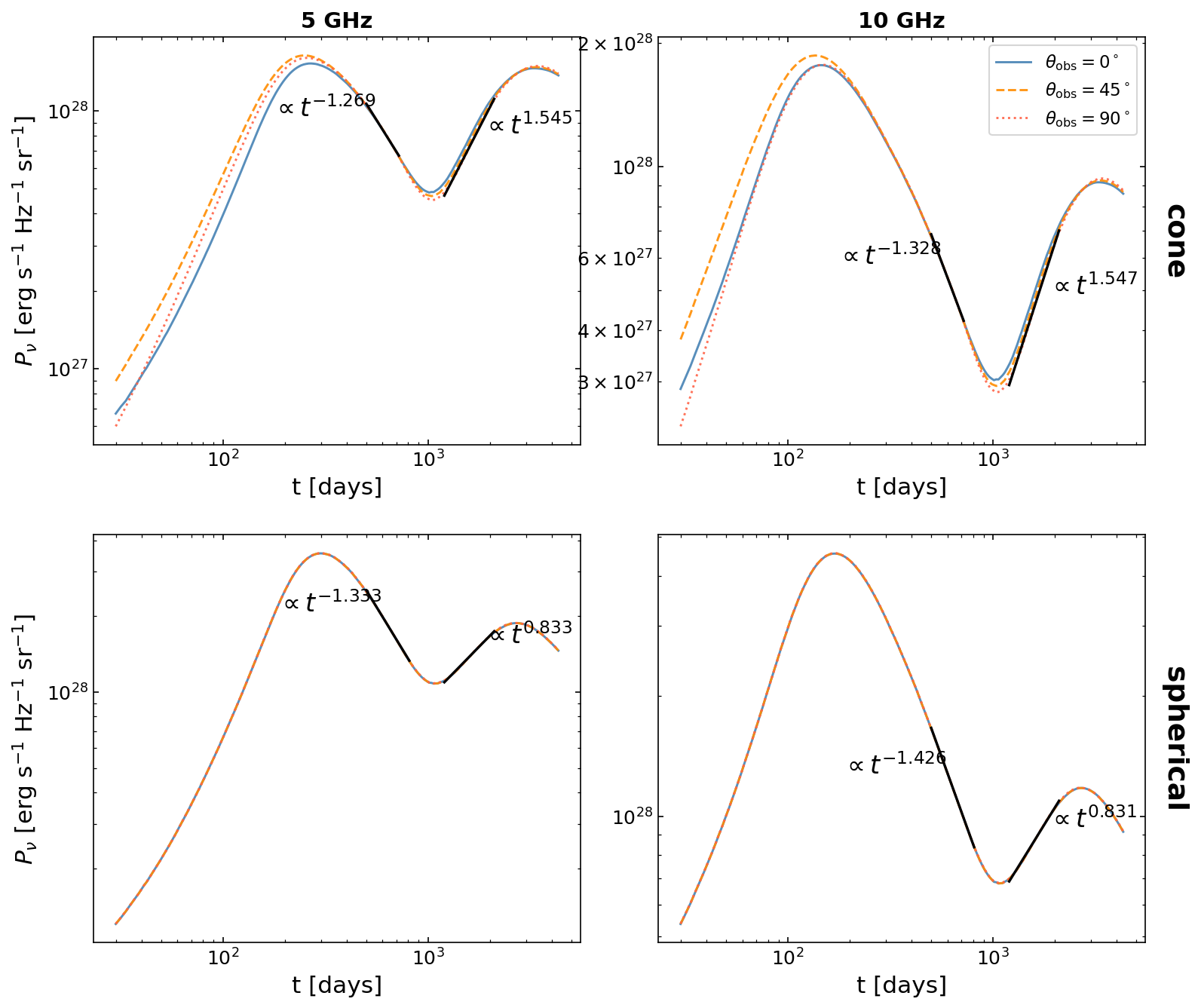}
    \caption{Light curves at 5 GHz (left column) and 10 GHz (right column) for the model Run-BP-cone (top row) and the model Run-BP-sph (bottom row). The blue solid, orange dashed, and red dotted lines indicate the light curves along the LoS at $\theta_{\rm obs} = 0^\circ$, $45^\circ$, and $90^\circ$, respectively.}
    \label{fig13}
\end{figure}

As proposed by \cite{2024ApJ...971...49M} and \cite{2026ApJ..1006..103L}, a broken power-law CNM profile — which decays rapidly inside the Bondi radius and becomes nearly flat beyond it — has the potential to explain the radio re-bursts observed in some sources. Initially, the emission at the observed frequency is optically thick and thus rises with time. If it becomes optically thin within the Bondi radius, the light curve begins to decay. After it crosses the Bondi radius, it rises again, producing a radio re-burst. Using a broken power-law CNM profile with $\rho \propto r^{-2.5}$ inside the Bondi radius and $\rho \propto r^0$ outside it, Fig.~\ref{fig13} shows the resulting double-peaked radio light curves for our two uniform, time-independent wind injection models.

The light curve of model Run-BP-cone resembles a combination of the two light curves shown in Fig.~\ref{fig12} for each frequency. The time of the first peak differs with frequency. The decline of the first peak follows $t^{-1.269}$ and $t^{-1.328}$ at 5 GHz and 10 GHz, respectively, which is slower than the $\sim -1.7$ decay seen in the single power-law simulations. This is because, at that time, the light curve has not yet fully entered the self-similar transition phase shown in Fig.~\ref{fig12}. At $t \sim 1000$~days, the wind crosses the Bondi radius, and the light curve begins to rise, with indices of $t^{-1.545}$ and $t^{-1.547}$ at 5 GHz and 10 GHz, respectively. These rise indices are also lower than those for a single power-law CNM profile. This is due to the fact that the CNM profile used in Fig.~\ref{fig12} sets the number density at the Bondi radius to $90~{\rm cm}^{-3}$, the same as in the broken power-law situation. Therefore, the CNM density inside the Bondi radius is higher for the broken power-law profile ($n = 90 \times (R_{\rm B}/r)^{2.5}$) than for a constant CNM profile, leading to stronger deceleration of the wind. The deceleration effect is not significant at the first peak because the light curve has not yet evolved to its final state and remains in the transition phase. At late times, the stronger deceleration affects the second peak, making its rise slower than in the single power-law case. For model Run-BP-sph, the decline of the first peak is faster than that of model Run-BP-cone due to the $\theta$ expansion of model Run-BP-cone, as described above, while the rise of the second peak is even slower (see Fig.~\ref{fig13}). This is due to the stronger deceleration in model Run-BP-sph, which sweeps up the CNM mass faster because its solid angle is significantly larger than that of Run-BP-cone, and the light curve power-law index is more sensitive to the deceleration for a flatter CNM profile (e.g., a smaller $b$, as described above).

\begin{figure}
    \centering
    \includegraphics[width=1.0\linewidth]{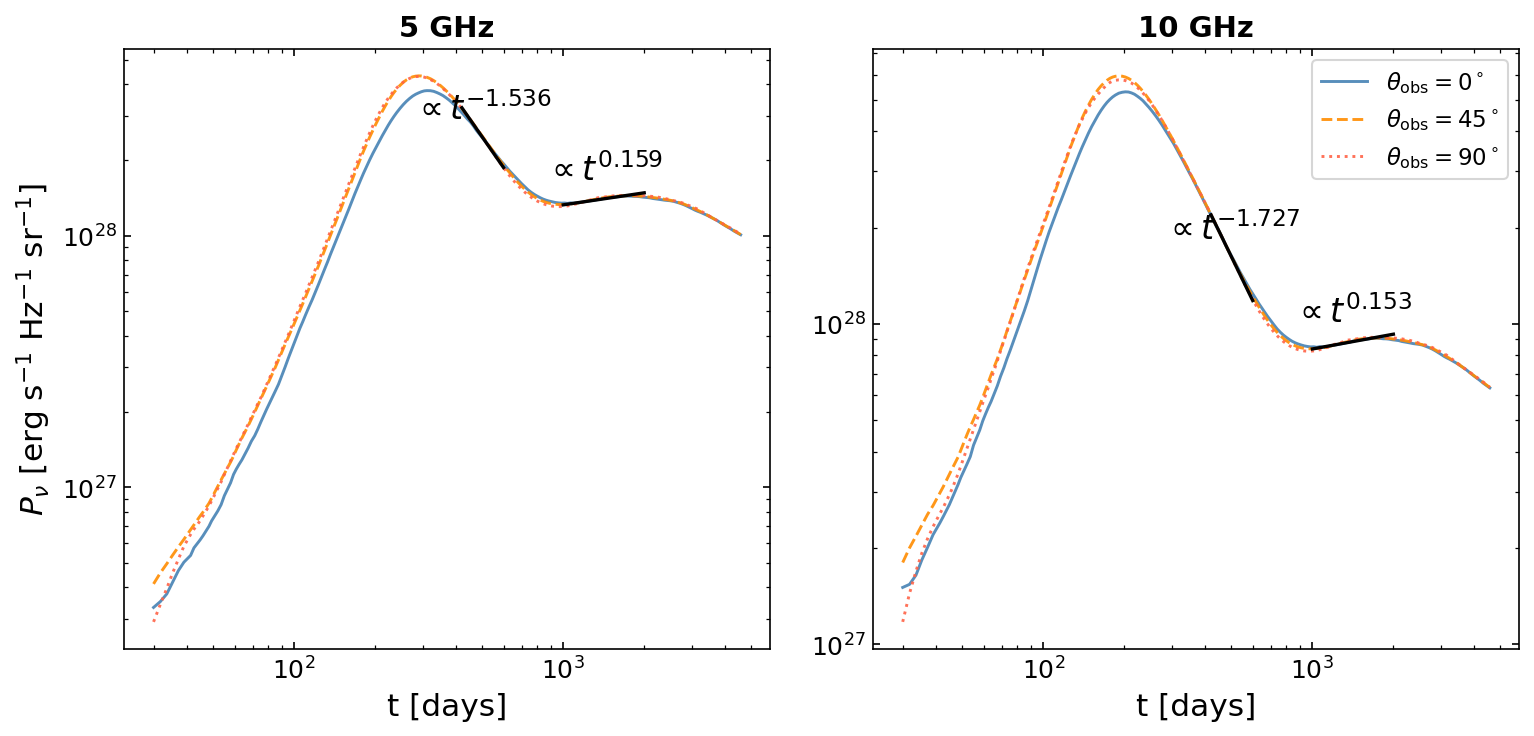}
    \caption{Light curves of the model Run-BP-M7 at 5 GHz (left panel) and 10 GHz (right panel). The format is the same as in previous figures.}
    \label{fig14}
\end{figure}

Finally, Fig.~\ref{fig14} shows the light curves of our model Run-BP-M7 at 5 GHz and 10 GHz. The angle-dependent deceleration effect is also evident here. At early times, while the uniform injection models decelerate slowly, for the model Run-BP-M7, the wind near the rotational axis — where the velocity is higher and the density is lower — decelerate more rapidly. As a result, the light curve decays faster (or rises more slowly) than in the uniform injection models. The decline of the first peak follows $t^{-1.536}$ and $t^{-1.727}$ at 5 GHz and 10 GHz, respectively, both of which are steeper than in the uniform wind injection case. The index at 10 GHz is lower (i.e., the decay is faster) than that at 5 GHz, because the 10 GHz light curve begins to decay earlier; by the time the wind crosses the Bondi radius, it has more fully entered into the self-similar decay phase than the 5 GHz light curve.
\begin{figure}
    \centering
    \includegraphics[width=1.0\linewidth]{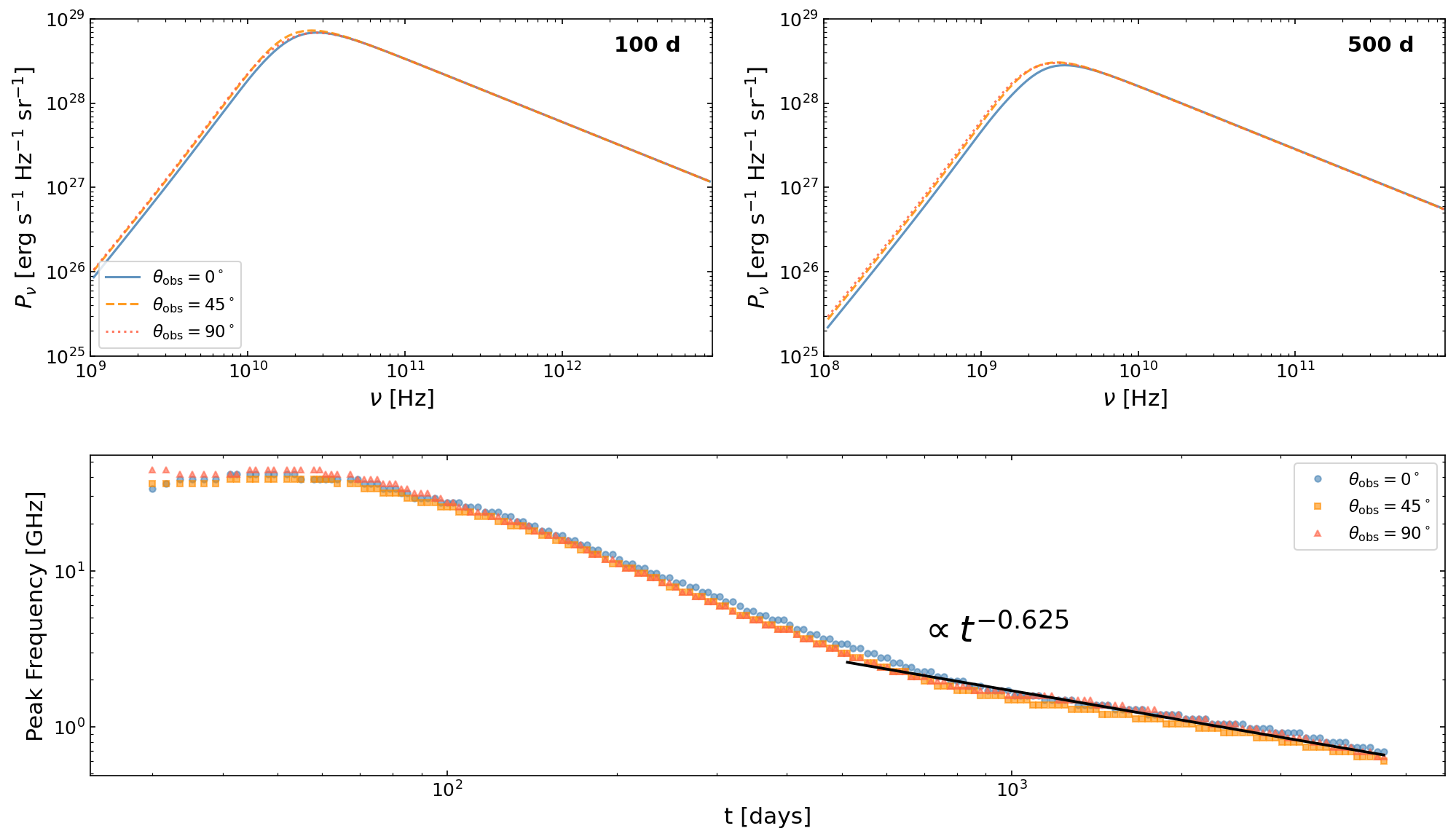}
    \caption{The top panels show the SED of the model Run-BP-M7 at $t = 100$~days (left) and $t = 500$~days (right). The bottom panel shows the temporal evolution of the peak frequency $\nu_{\rm p}$. The format is the same as in previous figures.}
    \label{fig15}
\end{figure}

While the light curve of model Run-SP-M7 at 10 GHz shows the effect of the inner shock component around $t \sim 200$~days (see Fig.~\ref{fig10}), the light curve of model Run-BP-M7 shows no sign of the inner shock. Fig.~\ref{fig15} shows the SED of Run-BP-M7, which is a typical single-component double power-law one. Its peak frequency is $\sim 30$ GHz at early times, which is much higher than that of Run-SP-M7. This is because the CNM density at small radii is much larger in the $r^{-2.5}$ regime than that in the $r^{-1}$ regime. This results in far more injected CRe in Run-BP-M7 than in Run-SP-M7 at early times. The forward shock component in this model is stronger than the inner shock component and dominates both the SED and the light curve. The effect of the inner shock is thus overwhelmed and cannot be seen in this situation.

In the model Run-SP-M7, the angular dependence of the deceleration causes the rise of the optically thin light curve to become a flat plateau (see Fig.~\ref{fig10}). The same phenomenon is also seen in the second peak for the broken power-law CNM profile. In the model Run-BP-M7, the rise of the second peak follows $t^{0.159}$ and $t^{0.153}$ for 5 GHz and 10 GHz, respectively, which is much slower than in the uniform injection models. The reason is the same as that described above.

Our simulations show that, even in the phase where the swept-up CNM mass is still less than the wind mass, a slow deceleration of wind occurs and affects the final light curve. Given that the kinetic energy and mass of wind, and CNM profile fixed, the deceleration of wind will be more obvious for the longer duration or lower density of wind. Consequently, for a more realistic TDE wind, in which both the density and radial velocity vary with angle, the deceleration process differs across different angles. Along each angle, the evolution of the shock velocity is distinct from that along other angles. Therefore, the sum of their contributions — the overall light curve — can differ significantly from analytical predictions. In our simulations, this results in a slower rise or a faster decline (depending on the CNM profile) than the analytical prediction, and can even turn the rise phase into a long-lasting plateau. Although our simulations only covers a limited set of parameters, and the models Run-SP-M7 and Run-BP-M7 are based on a specific wind property that may not be representative of all TDEs, this effect should be robust and should be taken into account when analyzing wind--CNM interactions. For instance, our simulations suggest that, if one intends to use a broken power-law CNM profile to explain the fast radio re-bursts observed in some sources, a rise in the CNM density outside the Bondi radius, or a stronger and shorter-duration wind, would be required.

\section{CONCLUSION AND DISCUSSION}
TDE wind-CNM interaction is believed to contribute to the radio emission of non-jetted TDEs. The wind-CNM interaction model has been studied both analytically and numerically. In previous simulations (e.g., \cite{2026ApJ..1006...70M}; \cite{2025ApJ...988L..24H}; \cite{2026ApJ..1006..103L}), the TDE wind is set to be time and angle-independent. It is not clear to what extent the results deviate from those in the case in which TDE wind is both time-variable and angle-dependent. 

We perform numerical simulations to investigate the effects of wind properties on the radio light curve in the wind-CNM interaction model. In our simulations, the CRes are injected according to the shock conditions. We can follow the evolution of the CRes in the simulations, which is important for the synchrotron radiation calculation (\cite{2026ApJ..1006...70M}).  In some simulations, we inject time-steady and angle independent wind. We have also simulations with time-variable and angle-dependent wind from the TDE simulations of \cite{BU22}. Our main findings are as follows:

\begin{itemize}
    \item Given that the wind mass and kinetic energy fixed, the radio light curve in the optically thin waveband significantly depends on the time-variability and angle-dependence of wind properties. 

    \item In the period after wind injection ($t > t_{\rm inject}$), the radio emission mainly comes from the forward shock propagating in the CNM. Even in the phase where the swept-up CNM mass is still less than the wind mass, a slow deceleration of wind occurs and makes the final light curve deviate from the theoretical prediction significantly. This results in a slower rise or a faster decline (depending on the CNM profile) than the analytical prediction, and can even turn the rise phase into a long-lasting plateau.

\end{itemize}

The double peak phenomenon in the radio light curve is proposed to be due to the radial broken power-law distribution of the CNM. The second rise is due to that the radial profile of CNM outside the Bondi radius is much flatter than that inside the Bondi radius. However, our results demonstrate, in model Run-BP-M7, due to overall much more obvious wind deceleration, the second flare has a significantly slower rise with time compared to the steady and angle-independent wind injection models. Thus, if one uses a broken power-law CNM profile to explain the fast radio re-bursts observed in some sources, the obtained radial profile of the CNM outside the Bondi radius may significantly depend on the angular settings of the properties of wind.

In summary, the results in this paper demonstrate that the radio light curve significantly depends on the angular-dependence of the properties of wind. It is essential to carefully take into account the angular dependence of wind properties to study the wind-CNM interaction.

In the wind-injection period, the time and angle-variability of wind results in many internal shocks (see model Run-SP-M7), which cause CRes injection. However, for steady wind injection, the internal shocks are not present. Thus, in this wind-injection period, the time-rising behavior of the optically thin waveband differs significantly between steady-wind injection model and time and angle-variable wind injection model. In the model with time-variable and angle-dependent wind injection (model Run-SP-M7), the optically thin radio emission flux increases significantly faster than that in the steady wind injection model. However, in internal shock regions, a fluid element may cross multiple shocks and be accelerated multiple times. The value of $\epsilon_{\rm e}$ may differ for electrons experiencing different acceleration episodes, and our current understanding of such multi-acceleration processes is limited. In future, it is important to study in significant details about the internal shocks to carefully estimate their contributions on the radio emission. 

\section{Disclosure of Delegation to Generative AI}
The authors declare the use of generative AI in the research and writing process. According to the GAIDeT taxonomy (2025), the following tasks were delegated to GAI tools under full human supervision:

- Code optimization
- Visualization

The GAI tool used was: ChatGPT-5.6.
Responsibility for the final manuscript lies entirely with the authors.
GAI tools are not listed as authors and do not bear responsibility for the final outcomes.
Declaration submitted by: Yu-Heng Sheng

\begin{acknowledgments}
This work is supported by the National Science and Technology Major Project of China (No. 2024ZD1100601). D. Bu is supported by the  Natural Science Foundation of China (grants 12673024, 12192220, 12192223) and the National SKA Program of China (No.2025SKA0130100). L.C. acknowledges support by the National SKA Program of China (Grant No.2022SKA0120102), the National Key R\&D program of China under the grant 2024YFA1611403, 2024YFA1611401, 2024YFA1611402, 2024YFA1611404 and Shanghai Pilot Program for Basic Research, Chinese Academy of Science, Shanghai Branch (JCYJ-SHFY-2021-013).
    
\end{acknowledgments}

\bibliography{sample701}{}
\bibliographystyle{aasjournalv7}



\end{document}